\documentclass[a4paper]{article}
\usepackage{geometry}

\usepackage{setspace}
\usepackage{float}
\usepackage{graphicx}
\usepackage{epstopdf}
\usepackage{hyperref}
\usepackage{standalone}
\usepackage{amsmath,amsfonts,amsthm,bm} 
\usepackage{color}

\usepackage{pdflscape}
\usepackage{authblk}
\usepackage{subcaption}
\usepackage[flushleft]{threeparttable}
\usepackage[comma,authoryear]{natbib}
\usepackage{enumitem}
\usepackage{xcolor}
\usepackage{soul}
\usepackage{makecell,booktabs}
\usepackage{caption,dcolumn}
\usepackage{verbatim} 
\newcolumntype{d}[1]{D{.}{.}{4}}

\usepackage{dsfont}

\usepackage{multirow}
\usepackage{array}
\newcolumntype{P}[1]{>{\centering\arraybackslash}p{#1}}

\usepackage [english]{babel}
\usepackage [autostyle, english = american]{csquotes}
\MakeOuterQuote{"}

\newcommand{\mbR}{{\mathbb R}}

\usepackage{mathtools}

\newcommand{\interior}[1]{%
 {\kern0pt#1}^{\mathrm{o}}%
}
\usepackage [english]{babel}
\usepackage [autostyle, english = american]{csquotes}
\usepackage{csquotes}

\usepackage{algorithm,algpseudocode}
\usepackage{caption}
\usepackage[makeroom]{cancel}

\makeatletter
\newcommand*\bigcdot{\mathpalette\bigcdot@{.5}}
\newcommand*\bigcdot@[2]{\mathbin{\vcenter{\hbox{\scalebox{#2}{$\m@th#1\bullet$}}}}}
\makeatother

\newtheorem{prop}{Proposition}

\newtheorem{Rem}{Remark}

\usepackage{xr}
\usepackage{tabularx}

\title{\textbf{Fast Computer Model Calibration using Annealed and Transformed Variational Inference}}

\author[1]{Dongkyu Derek Cho}
\author[2]{Won Chang}
\author[3,4]{Jaewoo Park}

\affil[1]{Department of Statistical Science, Duke University}
\affil[2]{Division of Statistics and Data Science,  University of Cincinnati}
\affil[3]{Department of Applied Statistics, Yonsei University}
\affil[4]{Department of Statistics and Data Science, Yonsei University}

\usepackage{pdfpages} 
\usepackage{pgffor} 

\makeatletter
\AtBeginDocument{\let\LS@rot\@undefined}
\makeatother

\def\supplementfilename{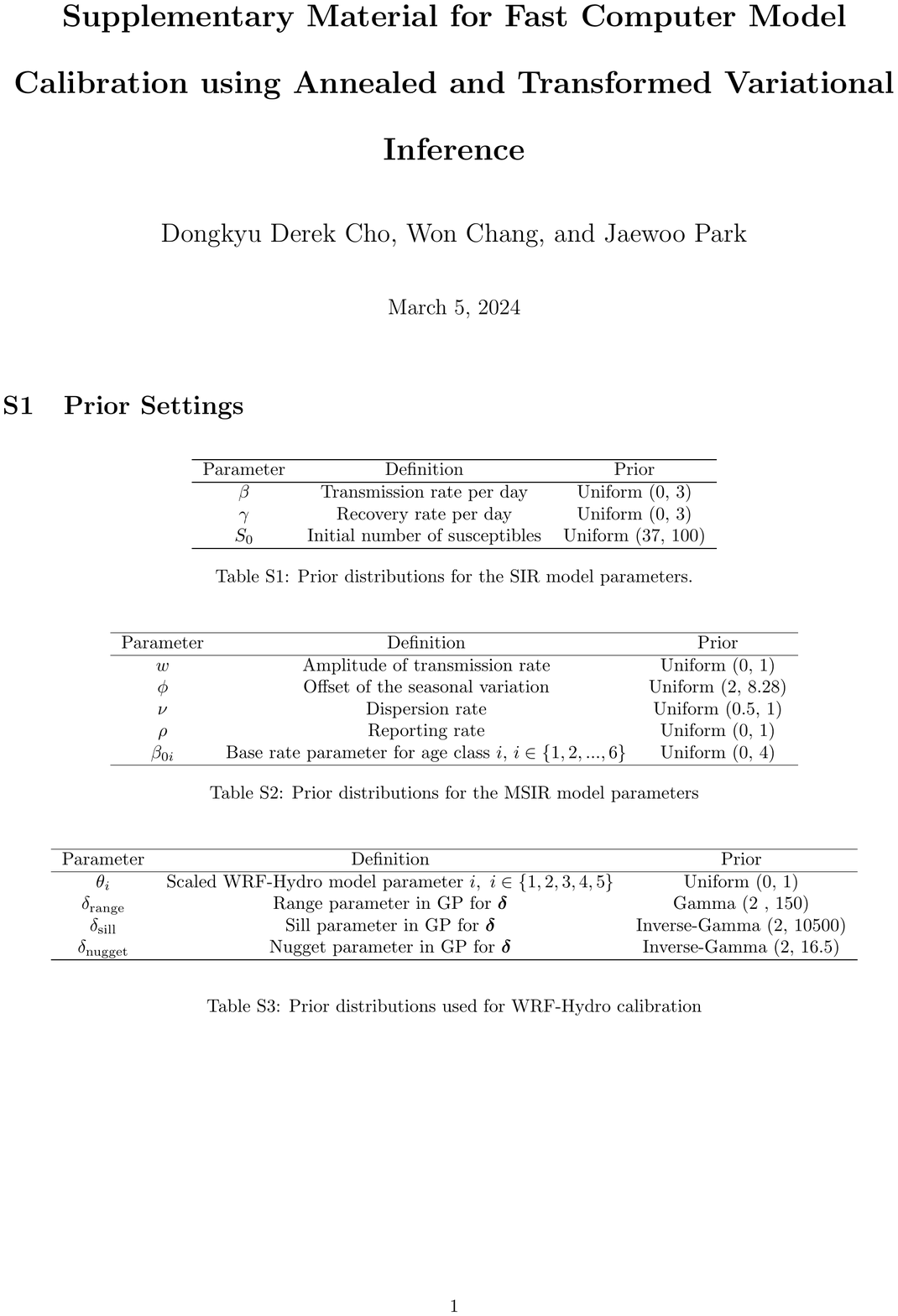}

\pdfximage{\supplementfilename}
\def\numbersupplementpages{\the\pdflastximagepages}

\newif\ifarXiv
\arXivtrue 

\begin{document}

\maketitle

\begin{abstract}
Computer models play a crucial role in numerous scientific and engineering domains. To ensure the accuracy of simulations, it is essential to properly calibrate the input parameters of these models through statistical inference. While Bayesian inference is the standard approach for this task, employing Markov Chain Monte Carlo methods often encounters computational hurdles due to the costly evaluation of likelihood functions and slow mixing rates. Although variational inference (VI) can be a fast alternative to traditional Bayesian approaches, VI has limited applicability due to boundary issues and local optima problems. To address these challenges, we propose flexible VI methods based on deep generative models that do not require parametric assumptions on the variational distribution. We embed a surjective transformation in our framework to avoid posterior truncation at the boundary. Additionally, we provide theoretical conditions that guarantee the success of the algorithm. Furthermore, our temperature annealing scheme can prevent being trapped in local optima through a series of intermediate posteriors. We apply our method to infectious disease models and a geophysical model, illustrating that the proposed method can provide fast and accurate inference compared to its competitors.
\end{abstract}

\noindent%


\section{Introduction}

Computer models have been widely used to simulate complex processes in many disciplines, including engineering, climate science, and epidemiology. One important source of uncertainty in utilizing a computer model is the parametric uncertainty due to the lack of or incomplete knowledge on input parameters that govern the behavior of the model. Computer model calibration, a formal statistical procedure to infer unknown input parameters, is crucial for generating realistic simulations \citep[e.g.,][]{kennedy2001bayesian,bayarri2007computer,Higdon2008,tuo2015efficient,plumlee2017bayesian,plumlee2019computer}. 
Although Bayesian calibration approaches have been developed and widely used, they can be computationally burdensome for complex models, because obtaining a well-mixed chain often requires a large number of iterations. In this manuscript, we propose a fast variational inference (VI) approach for calibrating computer models. By utilizing the deep generative methods, this new algorithm does not rely on a traditional class of variational distribution (e.g. Gaussian), allowing for a highly flexible posterior distribution.
Furthermore, we adopt surjective transformation and temperature annealing to avoid inferential challenges in a naive VI approach.

Compared to the simulation-based Bayesian algorithms such as Markov chain Monte Carlo (MCMC) or approximate Bayesian computation \citep{beaumont2002approximate, marin2012approximate}, VI approaches \citep{blei2017variational, kucukelbir2017automatic} can quickly approximate complex posteriors through optimization. However, even with such computational gains, VI is of limited applicability in computer model calibration for the following reasons. First, the posterior densities of input parameters in computer models are often highly irregular. The traditional VI based on strong assumptions on the variational distribution (e.g., the conditionally conjugate exponential family model and mean field approximations) may not be able to approximate such complex densities. Moreover, VI approaches often suffer from poor local optima problems \citep{blei2017variational,zhang2018advances} for highly complicated posterior densities. Second, it is often required to impose range constraints on the parameters in computer models, as these parameters are subject to some scientific constraints. Although bijective transformations that map from constrained to unconstrained parameter space have been widely used, such transformations may result in undesirable boundary effects at the edge of the parameter space \citep{kucukelbir2017automatic}. 


To address such challenges, we propose a VI algorithm for calibrating computer models. Our annealed and transformed VI (ATVI) combines key ideas from surjective transformations and temperature annealing. We adopt a surjective function \citep{nielsen2020survae} that maps unconstrained to constrained parameter space to avoid boundary effects caused by bijective transformations. We provide theoretical conditions that guarantee the success of surjective transformation and suggest a practical choice for such transformation. Furthermore, we embed the sequence of intermediate posteriors in the normalizing flow framework to avoid local optima issues, motivated by the temperature annealing \citep{neal2001annealed,mandt2016variational}. 

There is growing literature on the deep generative models in the deep learning community. Deep generative models have shown usefulness in many tasks as they are easily scalable and can represent high-dimensional correlated structures (see \cite{bond2021deep} for a review). We propose a practical tool for calibrating computer models by combining the deep generative model and VI frameworks. To the best of our knowledge, this is the first attempt to study deep learning-based VI for calibrating computer models. Through multiple challenging computer model examples, we observe that our method is faster than the traditional Bayes approaches (e.g., MCMC) while accurate compared to a na\"ive VI. Moreover, in some cases, our method can outperform the standard MCMC method by better handling the local maxima issues.

The outline for this paper is as follows. In Section 2, we introduce the background for computer model calibration. In Section 3, we discuss the application of deep learning-based VI to computer models and the inherent challenges. In Section 4, we propose ATVI by adapting surjective transformation and embedding temperature annealing schemes. In Section 5, we study the performance of ATVI through simulated and real data examples. We conclude with a discussion in Section 6. 

\section{Background}

\subsection{Bayesian Calibration}

While the computer models provide a helpful way to model the dynamics of real world phenomena, they are subject to various sources of uncertainties surrounding the model. One important source of uncertainty is the parametric uncertainty, stemming from the fact that the proper values of the parameters in the computer model are not known \emph{a priori} and hence need to be estimated based on the data. The statistical procedure for estimating these parameters are referred to as model calibration, and there has been a suite of studies on this topic in the literature  \citep[e.g.][]{kennedy2001bayesian,bayarri2007computer,Higdon2008,chang2015binary,tuo2015efficient,plumlee2017bayesian,plumlee2019computer,bhatnagar2022computer} over the recent decades.

In a calibration problem, our goal is to estimate the input parameter $\bm{\theta} \in \bm{\Theta} \subset \mbR^{d}$ based on the observed data $\mathbf{D} \in \mbR^{p}$. Here, $d$ denotes the number of input parameters to be calibrated. The observed data are typically in the form of time series \citep[e.g.,][]{bayarri2007computer,Higdon2008,bhatnagar2022computer} or spatial data \cite[e.g.]{chang2013fast,chang2015binary}, and therefore $p$ is determined as the length of observed time series or the number of observed locations. The general calibration framework \citep{kennedy2001bayesian} can be written as follows:
\vspace{-10pt}
\begin{equation} \label{eqn:basic_model}
\mathbf{D} = A(\bm{\theta}) + \bm{\delta}.
\vspace{-10pt}
\end{equation} 
where $A(\bm{\theta})$ is the outcome of the computer model with the same format as $\mathbf{D}$ at the input parameter setting $\bm{\theta}$ and $\bm{\delta}$ is the data-model discrepancy. Depending on the problem, the discrepany term $\bm{\delta}$ can be assumed to be a simple noise with an i.i.d. distribution or a complicated errors with temporal or spatial dependence. 


Let $p(\mathbf{D}|\bm{\theta})$ be the likelihood function based on the model in \eqref{eqn:basic_model} and $p_{\bm{\theta}}(\bm{\theta})$ be the prior density of $\bm{\theta}$. Then we can construct the posterior density $p_{\bm{\theta}}(\bm{\theta}|\mathbf{D}) \propto p_{\bm{\theta}}(\bm{\theta})p(\mathbf{D}|\bm{\theta})$ for the Bayesian inference. MCMC is a standard way to infer $\bm{\theta}$ in such a situation. However, inference based on MCMC can be ineffective because the computer model $A(\bm{\theta})$ is a highly complex function of the input parameter $\bm{\theta}$ and hence the posterior density $p_{\bm{\theta}}(\bm{\theta}|\mathbf{D})$ has a complicated form with multiple modes or strong dependence among the elements in $\bm{\theta}$. 
In such a situation, obtaining a well-mixed chain can take several hours or more to fully recover the posterior density, making the MCMC-based Bayesian method unattractive to practitioners who often need to provide projections in a timely manner.

\section{Variational Inference Using Normalizing Flows}

\subsection{Variational Inference with Parameter Constraints}

VI approximates the posterior distribution through optimization instead of computationally expensive sampling procedures such as MCMC \citep{blei2017variational, kucukelbir2017automatic}. The main idea is to approximate the target posterior density $p_{\bm{\theta}}(\bm{\theta}|\mathbf{D})$ using a variational density $q_{\bm{\theta}}(\bm{\theta};\bm{\phi})$ where  $\bm{\phi}$ is the variational parameter. With respect to $\bm{\phi}$, we minimize the reverse Kullback-Leibler (KL) divergence between the variational density and the posterior as follows:  
\vspace{-8pt}
   \begin{align}
    \text{KL}(q_{\bm{\theta}}(\bm{\theta};\bm{\phi})||p_{\bm{\theta}}(\bm{\theta}|\mathbf{D})) = -\mathbb{E}_{ q_{\bm{\theta}}(\bm{\theta};\bm{\phi})}[{\log p(\mathbf{D}| \bm{\theta}) + \log p_{\bm{\theta}}(\bm{\theta}) - \log p(\mathbf{D}) - \log q_{\bm{\theta}}(\bm{\theta};\bm{\phi})}]. \label{VIeq0} 
 \end{align}
By ignoring the normalizing constant $p(\mathbf{D})$ that is not relevant to the target parameter $\bm{\theta}$, we can obtain the evidence lower bound (ELBO) as 
\vspace{-8pt}
\begin{align} 
\mathcal{L} = \mathbb{E}_{ q_{\bm{\theta}}(\bm{\theta};\bm{\phi})}[{\log p(\mathbf{D}| \bm{\theta}) + \log p_{\bm{\theta}}(\bm{\theta}) - \log q_{\bm{\theta}}(\bm{\theta};\bm{\phi})}] \label{VIeq1}.
\end{align}
Then we maximize ELBO in \eqref{VIeq1}, which is equivalent to minimizing the KL divergence in \eqref{VIeq0}.
However, the ELBO in \eqref{VIeq1} cannot be directly applicable to computer model calibrations because of parameter constraints. In many computer models, parameters are limited within certain ranges due to scientific constraints. 

To address this, we introduce a bijective transformation $\mathcal{T} : \bm{\Xi} \rightarrow \bm{\Theta}$, where $\bm{\Theta}$ and $\bm{\Xi}$ are the constrained and unconstrained parameter space, respectively \citep{kucukelbir2017automatic}. We denote the $i$th element of $\bm{\theta} = (\theta_{1}, \theta_{2}, ..., \theta_{d})$ as $\theta_{i}$, and the lower and upper bounds of $\theta_{i}$  as $a_i$ and $b_i$  respectively. The constrained space can be represented as a Cartesian product of the compact spaces, $\bm \Theta = \prod_{i=1}^{d}[a_i, b_i]$. Then the inverse mapping is   $\mathcal{T}^{-1}(\bm{\theta})=\bm{\xi}$, where $\bm{\xi} \in \bm{\Xi} = \prod_{i=1}^{d} (-\infty, \infty)$. With such variable transformation, we can represent the variational density as
\begin{equation} \label{transform1}
\log q_{\bm{\theta}}(\bm{\theta};\bm{\phi}) = \log q_{\bm \xi}(\bm \xi ;\bm{\phi}) + \log\left|\det \frac{\partial \mathcal{T}(\bm{\xi})}{\partial \bm{\xi}}\right|.
\end{equation} 
By plugging \eqref{transform1} into \eqref{VIeq1}, we can represent our objective function  \eqref{VIeq1} as
\begin{equation}
\label{ELBO_trans}
\begin{split}
\mathcal{L} &= \mathbb{E}_{ q_{\bm{\theta}}(\bm{\theta};\bm{\phi})}[{\log p(\mathbf{D}| \bm{\theta}) + \log p_{\bm{\theta}}(\bm{\theta}) - \log q_{\bm{\theta}}(\bm{\theta};\bm{\phi})}] \\
&= \mathbb{E}_{ q_{\bm{\theta}}(\bm{\theta};\bm{\phi})} \left[\log p(\mathbf{D}| \bm{\theta}) + \log p_{\bm{\theta}}(\bm{\theta}) - \log q_{\bm \xi}(\bm \xi ;\bm{\phi}) - \log\left|\det \frac{\partial \mathcal{T}(\bm{\xi})}{\partial \bm{\xi}}\right| \right].
\end{split}
\end{equation}
Then we can maximize \eqref{ELBO_trans} with respect to $\bm{\phi}$, while considering the parameter constraints in $\bm{\theta}$.

\subsection{Deep Approximate Distribution with Normalizing Flow} 

To approximate the variational density  $q_{\bm{\theta}}(\bm{\theta};\bm{\phi})$, we utilize normalizing flows (NF) \citep{papamakarios2021normalizing, kobyzev2020normalizing} that can learn a non-linear and invertible mapping from a simple latent distribution to a target distribution. The approximate distribution given by normalizing flow can be viewed as a generalization of the reparameterization trick \citep{papamakarios2021normalizing}. If we have a well-trained NF model $f$, we can simply sample a random vector $\mathbf{z}_{0} \in \mathbb R^{d}$ from a latent distribution with the density $p_{\mathbf{z}_{0}}(\mathbf{z}_{0})$; especially, we use the standard normal density for $\mathbf{z}_{0}$. Then we can supply $\mathbf{z}_{0}$ to the NF model to obtain $\bm{\xi} = f(\mathbf{z}_{0} ;\bm{\phi})$, which is a sample from our approximate distribution with the density $q_{\bm \xi}(\bm \xi ;\bm{\phi})$. From now on, we refer to the variational density as the approximate density as well. With the variable transformation, the approximate density can be represented as
\begin{equation}
     \log q_{\bm \xi}(\bm \xi ;\bm{\phi}) = \log (p_{\mathbf{z}_{0}}(\mathbf{z}_{0})) + \log \left|\det \frac{\partial f(\mathbf{z}_{0};\bm{\phi})}{\partial \mathbf{z}_{0}} \right| \label{NF_formula}.
\end{equation}

The NF model is defined with a stack of invertible and differentiable functions. Starting from the initial distribution for $\mathbf{z}_0$, the approximate distribution gradually evolves through a series of computationally efficient transformations. As a result, the model can represent a highly complex approximate distribution with relatively low computational cost while utilizing all the computational and inferential tools for modern deep learning. Compared to other traditional methods such as mean-field approximation \citep{zhang2018advances}, the NF model does not have to rely on specific parametric forms of variational distribution.

Furthermore, the NF model has some significant advantages when used for VI. The NF model exploits a reparameterization trick, which represents the gradient estimate as a series of deterministic mapping of a latent distribution. This leads to a lower variance for the resulting estimator than the other stochastic gradient estimators \citep{kingma2013auto,zhang2018advances}. Unlike other deep generative models, the NF model provides a way to explicitly calculate the density $q_{\bm{\theta}}(\bm{\theta};\bm{\phi})$, which removes the need for approximating the $\log q_{\bm{\theta}}(\bm{\theta};\bm{\phi})$ in \eqref{VIeq1}. Due to such advantages, NF models have been widely used for variational inference \citep{rezende2015variational,kingma2016improved,berg2018sylvester}. However, to our knowledge, this study is the first attempt to utilize NF models for computer model calibration. 

Among the model specifications available in the literature, we use a stack of rational-quadratic neural spline layers in an autoregressive way to construct the NF model $f$. The theoretical background of this function is given by \cite{gregory1982piecewise}, and its usefulness in normalizing flow was demonstrated by \cite{durkan2019neural}. For illustration, consider a single autoregressive rational quadratic spline layer $f = \{f^{(i)}\}_{i=1}^{d_{\mathbf{x}}}$ with an input $\mathbf{x} \in \mathbb{R}^{d_{\mathbf{x}}}$ and an output $\mathbf{y} \in \mathbb{R}^{d_{\mathbf{x}}}$. 
The parameterization function takes input
$\mathbf{x}^{(i)} = [\mathbf{x}_{1},...,\mathbf{x}_{i-1}]$ and outputs $\mathbf{h}^{(i)} \in \mathbb{R}^{d_{3K-1}}$, the parameters of $K$ spline functions. Outputs are divided into three parts, $\mathbf{h}^{(i)} = [\mathbf{h}^{(i),\textbf{w}}, \mathbf{h}^{(i),\textbf{e}}, \mathbf{h}^{(i),\bm{\rho}}]$ with length $K, K, K-1$, respectively. These outputs are transformed through softmax or softplus functions and rescaled. The overall structure can be summarized as
\begin{align}
\begin{split} \label{equation:NFoverall}
      &\mathbf{h}^{(i)} = [\mathbf{h}^{(i),\textbf{w}}, \mathbf{h}^{(i),\textbf{e}}, \mathbf{h}^{(i),\bm{\rho}}] = \sigma_{i}(\mathbf{x}^{(i-1)}), \\
      &\textbf{w}^{(i)} = 4  \textit{Softmax}(\mathbf{h}^{(i),\textbf{w}}), \\ 
      &\textbf{e}^{(i)} = 4  \textit{Softmax}(\mathbf{h}^{(i),\textbf{e}}), \\
    &\bm{\rho}^{(i)} = \textit{Softplus}(\mathbf{h}^{(i), \bm{\rho}}),  
\end{split}
\end{align}
where $\sigma_{i}: \mathbb{R}^{i-1} \rightarrow \mathbb{R}^{3K-1}$ can be any flexible transformation function. In our implementation, we use the standard multilayer perceptron model for all $\sigma_{i}$. The range for each output is given by $\textbf{w}^{(i)}, \textbf{e}^{(i)} \in [-2, 2]^{K}$ and $\bm{\rho}^{(i)} \in [0, 1]^{K-1}$. The terms $\textbf{w}^{(i)}$, $\textbf{e}^{(i)}$, and $\bm{\rho}^{(i)}$ in \eqref{equation:NFoverall} are used to build the knot locations and the spline basis. We provide more details in Section S2.2 in the Supplementary Material.

\subsection{Challenges in Variational Inference}
\label{section:challenges}

 Compared to the standard MCMC algorithms, VI approaches scale well to larger datasets or high-dimensional problems. Furthermore, with the well-trained NF model, we can construct a complicated form of variational distribution without depending on restrictive parametric assumptions. However, the direct application of these methods to computer model calibration is challenging due to the following reasons. 

The first issue is local optima. If the approximate density $q_{\bm{\theta}}(\bm{\theta};\bm{\phi})$ does not cover (i.e., mismatch) all of the high-density regions of the target $p_{\bm{\theta}}(\bm{\theta} | \mathbf{D})$ at the beginning of the training, the mismatched region may never be discovered during optimization \citep{regli2018alpha,turner+sahani:2011a}. Although several alternatives, such as the trust region update \citep{arenz2020trust} or employing different divergences \citep{ranganath2016operator} have been proposed, they are computationally expensive and require some simplifying assumptions in the form of variational density. Our proposed solution is to start optimization with a well-constructed $q_{\bm{\theta}}(\bm{\theta};\bm{\phi})$ which does not have too much mismatch with $p_{\bm{\theta}}(\bm{\theta} | \mathbf{D})$. We provide more details with a practical solution in Section 4.3.

Furthermore, the standard VI suffers from boundary effect problems due to parameter constraints in computer models. 
For illustration, consider a parameter $\theta \in \bm{\Theta} = [0,1]$ with a commonly used  logistic function $\mathcal{T}$, which maps $\xi \in \bm{\Xi}$ to $\theta \in \bm{\Theta}$ (i.e., $\theta = \mathcal{T}(\xi) = 1/(1+ e^{-\xi})$). Then with a variable transformation, we have
\vspace{-10pt}
\begin{align}
    q_{\theta}(\theta ; \bm{\phi}) &= q_{\xi}(\xi ; \bm{\phi})\left|\text{det}\frac{\partial \mathcal{T}(\xi)}{\partial \xi}\right| \nonumber\\
    &= q_{\xi}(\xi ; \bm{\phi}) \left|\mathcal{T}(\xi)(1 - \mathcal{T}(\xi)) \right| \label{transformbound}.
    \vspace{-5pt}
\end{align}
The representation in \eqref{transformbound} prevents $q_{\theta}(\theta ; \bm{\phi})$ from having positive values near the boundary of the $\theta$ (i.e., 0 or 1), because $|\mathcal{T}(\xi)(1 - \mathcal{T}(\xi))| \rightarrow 0$ as $\xi \rightarrow \pm \infty$. Other bijective transformations, such as hyperbolic tangent, can also have boundary effect issues. When the target density $p_{\theta}(\theta | \mathbf{D})$ has a positive mass at the boundary of $\bm{\Theta}$, the approximate density by $q_{\theta}(\theta ; \bm{\phi})$ will have a substantial mismatch with $p_{\theta}(\theta | \mathbf{D})$ near the boundary areas.

\section{Annealed and Transformed Variational Inference}

\begin{figure}[htbp]
\begin{center}
\includegraphics[width=0.55\columnwidth]{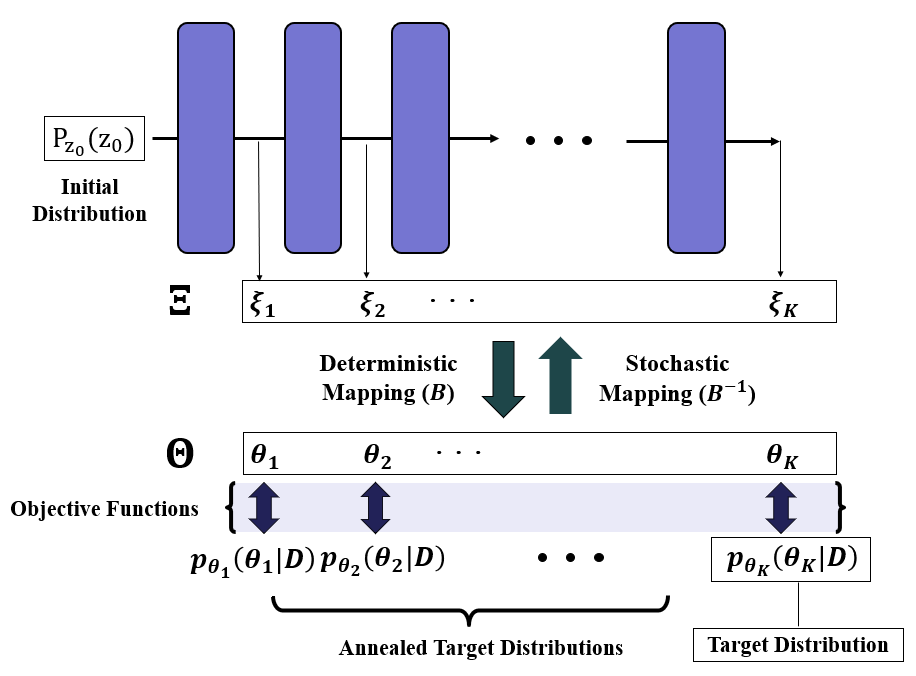}
\end{center}
\caption[]{Illustration for the ATVI.}
\label{Model_Outline}
\end{figure}

In this section, we propose an annealed and transformed VI (ATVI) method for computer models as a solution to the challenges described in Section \ref{section:challenges}. An outline of ATVI is described in Figure~\ref{Model_Outline}. Figure~\ref{Model_Outline} illustrates how the initial latent variable $\mathbf{z}_0$ is sequentially transformed into the target parameter $\bm{\theta}$. Here, we propose a new temperature annealing scheme for VI to construct intermediate target distributions, approximated with multiple real NVP layers. These block structures allow us to avoid local minima issues. In addition, we introduce a \emph{surjective} transformation $\mathcal{B}$ instead of the bijective transformation $\mathcal{T}$ to resolve the boundary effect issues.

\vspace{-15pt}
\subsection{Boundary Surjection}
\label{section:surjection}

To address parameter constraint issues, we propose a new transformation, the boundary surjection. The boundary surjection guarantees the parameter constraints using a simple transformation. A mapping $\mathcal{B}: \bm \Xi \rightarrow \bm \Theta$ is a deterministic and surjective mapping, which ensures that the mapping from $\bm \xi$ to $\bm \theta$ satisfies the parameter constraints. However,  $\mathcal{B}^{-1}: \bm \Theta \rightarrow \bm \Xi$ is not deterministic, and we refer to this inverse mapping as a stochastic transformation. 
 
While most NF models are strictly bijective, \cite{nielsen2020survae} suggested a stochastic and surjective transformation in their `SurVAE' model. In this framework, the transformation is evaluated through a \emph{likelihood contribution}, which works similarly to the Jacobian term in  \eqref{transform1}. \cite{nielsen2020survae} showed that for any distribution that is defined in $\bm \Xi$ or $\bm \Theta$, a surjective transformation can be approximated as follows: 
\vspace{-10pt}
\begin{align} \label{transform2}
      \mathbb{E}_{v_{\bm{\theta}}(\bm{\theta}|\bm{\xi})}\left[\log q_{\bm{\theta}}(\bm{\theta}; \bm{\phi})\right]  &\approx \log q_{\bm{\xi}}(\bm{\xi}; \bm{\phi}) - \overbrace{ \mathbb{E}_{ v_{\bm{\theta}}(\bm{\theta}|\bm{\xi})} \left[\log \frac{w_{\bm{\xi}}(\bm{\xi}|\bm{\theta})}{v_{\bm{\theta}}(\bm{\theta}|\bm{\xi})} \right]}^{\mathcal{V}(\bm{\theta}, \bm{\xi})},
\end{align}
where $v$ and $w$ are auxiliary probability densities. Here $\mathcal{V}(\bm{\theta}, \bm{\xi})$ is called the likelihood contribution which is always non-positive by definition. If the transformation is bijective, the likelihood contribution becomes deterministic and identical to the Jacobian term in  \eqref{transform1}. \cite{nielsen2020survae} used their SurVAE model to impose the symmetry, permutation-invariance, or low dimensionality condition in learning the distribution of images or spatial fields. On the other hand, our surjective transformation focuses on alleviating boundary effect issues in calibration problems. 


Consider the constrained parameter space $\bm \Theta = \prod_{i=1}^{d}[a_i, b_i]$, which is defined through the Cartesian product of the compact spaces. Here, for each $i$ we introduce a boundary surjection marginally through an augmented function $g$ defined as follows:
\vspace{-3pt}
\begin{equation}
\label{mappingdef}
\begin{split}
 g(\xi_{i})=(\theta_i, s_{i}) =\left \{
\begin{array}{l}(2b_{i} - \xi_{i}, 2), \ \text{if} \ \xi_{i} > b_{i},\\[1ex]
{}(\xi_{i} ,1),\ \text{if} \ a_{i}<\xi_{i}<b_{i},\\[2ex]
{}(2a_{i} - \xi_{i}, 0),\ \text{if} \ \xi_{i}<a_{i},\\ \end{array} \right.
\end{split}
\end{equation}
\vspace{-25pt}
\begin{equation}
\begin{split}
\xi_{i} = g^{-1}(\theta_i, s_{i})=\left \{
\begin{array}{l} 2b_{i} - \theta_{i}, \ \text{if} \ s_{i} = 2,\\[1ex]
{} \theta_{i},\ \text{if} \ s_{i}=1,\\[2ex]
{} 2a_{i} - \theta_{i},\ \text{if} \ s_{i}=0.\\ \end{array} \right.
\end{split}
\end{equation}
Then the boundary surjection is given as $\mathcal{B}(\xi_{i}) =g_{1}(\xi_{i})$, where $\theta_i=g_{1}(\xi_{i})$ and $s_i=g_{2}(\xi_{i})$ are the first and second components of the augmented function $g(\xi_{i})$, respectively. 
Note that the transformation $\mathcal{B}(\xi_{i})$ is both surjective and non-injective $\forall \xi_{i} \in \mathcal{B}^{-1}([a_{i}, b_{i}])$. The constrained parameter value $\theta_i$ is computed deterministically for a  given value of $\xi_i$, but a tuple $(\xi_{i}, s_{i})$ cannot be obtained deterministically when $\theta_{i}$ is given. 

The auxiliary probability density $v_{\theta_{i}}(\theta_{i}|\xi_{i})$ in \eqref{transform2} can be decomposed as 
\begin{equation} 
\label{auxilary_distribution}
\begin{split}
v_{\theta_{i}}(\theta_{i}|\xi_{i}) &= \sum_{s_{i} \in \{0,1,2\}}{v_{\theta_{i}}(\theta_{i}|\xi_{i},s_{i})}{v_{s_{i}}(s_{i}|\xi_{i})}.
\end{split}
\end{equation}
Once we set the densities $v_{\theta_{i}}(\theta_{i}|\xi_{i},s_{i})$ and $v_{s_{i}}(s_{i}|\xi_{i})$  to be Dirac delta functions whose values are zero everywhere except at $g_{1}(\xi_{i})$ and $g_{2}(\xi_{i})$ respectively, $v_{\theta_{i}}(\theta_{i}|\xi_{i})$ in \eqref{auxilary_distribution} becomes 
$\sum_{s_{i} \in \{0,1,2\}}{\delta(\theta_{i} - g_{1}(\xi_{i}))}{\delta(s_{i}-g_{2}(\xi_{i}))}$. In a similar fashion, we can represent the other auxiliary probability density $w_{\xi_{i}}(\xi_{i}|\theta_{i})$ as 
\begin{equation} 
\label{auxilary_distribution2}
\begin{split}
w_{\xi_{i}}(\xi_{i}|\theta_{i}) &= \sum_{s_{i} \in \{0,1,2\}}{w_{\xi_{i}}(\xi_{i}|\theta_{i},s_{i})}{w_{s_{i}}(s_{i}|\theta_{i})},
\end{split}
\end{equation}
where $w_{\xi_{i}}(\xi_{i}|\theta_{i}, s_{i})$ is a Dirac delta function whose values are zero everywhere except at $g^{-1}(\theta_i,s_{i})$ and $w_{s_{i}}(s_{i}|\theta_{i})$ is the conditional probability mass function for $s_{i}$ given $\theta_{i}$. Then \eqref{auxilary_distribution2} becomes $\sum_{s_{i} \in \{0,1,2\}}{\delta(\xi_{i} - g^{-1}(\theta_{i},s_{i}))}{w_{s_{i}}(s_{i}|\theta_{i})}$. Compared to  \eqref{auxilary_distribution}, \eqref{auxilary_distribution2} is a stochastic transformation due to the randomness in $s_{i}$; therefore, the inverse mapping $\mathcal{B}^{-1}$ is stochastic. From \eqref{auxilary_distribution} and \eqref{auxilary_distribution2}, the likelihood contribution for each $i$ can be simplified as
\begin{equation} 
\label{likelihood_contribution}
\begin{split}
\mathcal{V}(\theta_{i}, \xi_{i}) &= \mathbb{E}_{v_{\theta_{i}}(\theta_{i}|\xi_{i})} \left[\log \frac{w_{\xi_{i}}(\xi_{i}|\theta_{i})}{v_{\theta_{i}}(\theta_{i}|\xi_{i})} \right]\\
&= \mathbb{E}_{ \sum_{s_{i}} v_{\theta_{i}}(\theta_{i}|\xi_{i},s_{i})v_{s_{i}}(s_{i}|\xi_{i})} \left[ \sum_{s_{i}}\log \frac{w_{\xi_{i}}(\xi_{i}|\theta_{i},s_{i})w_{s_{i}}(s_{i}|\theta_{i})}{v_{\theta_{i}}(\theta_{i}|\xi_{i},s_{i})v_{s_{i}}(s_{i}|\xi_{i})}\right] \\
&= \mathbb{E}_{\delta(\theta_{i} - g_{1}(\xi_{i}))\delta(s_{i}-g_{2}(\xi_{i}))} \left[\log \frac{\delta(\xi_{i} - g^{-1}(\theta_{i},s_{i}))w_{s_{i}}(s_{i}|\theta_{i})}{\delta(\theta_{i} - g_{1}(\xi_{i}))\delta(s_{i}-g_{2}(\xi_{i}))} \right] \\
&= \log w_{s_{i}}(s_{i}|\theta_{i}).
\end{split}
\end{equation}
In \eqref{likelihood_contribution}, the third line results from the fact that for an observed $\xi_{i}$, $v_{s_{i}}(s_{i}|\xi_{i})$ takes a positive value only when $s_{i}=g_{2}(\xi_{i})$. Similarly, the last equality holds because we defined $g_{1}(\xi_{i})=\theta_{i}$ and $g_{2}(\xi_{i})=s_{i}$ in \eqref{mappingdef}. The joint likelihood contribution for all parameters can be written as $\mathcal{V}(\bm{\theta}, \bm{\xi}) = \sum_{i=1}^{d}\mathcal{V}(\theta_{i},\xi_{i})$. Similarly, we 
write the joint auxiliary densities as $v_{\bm{\theta}}(\bm{\theta}|\bm{\xi}) = \prod_{i=1}^{d} v_{\theta_{i}}(\theta_{i}|\xi_{i})$, and $w_{\bm{\xi}}(\bm{\xi}|\bm{\theta}) = \prod_{i=1}^{d} w_{\xi_{i}}(\xi_{i}|\theta_{i})$. Based on this simplified likelihood contribution, we summarize the transformed variational inference (TVI) algorithm in  Algorithm~\ref{algo_boundarysurjection}. 

\begin{algorithm}
\caption{Transformed Variational Inference (TVI) algorithm }\label{algo_boundarysurjection}
\begin{algorithmic}[H]
\small
\State 1: Initialize $\bm{\phi}$.

\State 2: \textbf{while} \textit{$\mathcal{L}$ not converged} \textbf{do}

\State 3: \indent  Sample $\{\mathbf{z}_{0}^{(j)}\}_{j=1}^{m} \sim p_{\mathbf{z}_{0}}(\mathbf{z}_{0})$ and get $\{\bm{\xi}^{(j)}\}_{j=1}^{m}$ using $\bm{\xi}^{(j)} = f(\mathbf{z}_{0}^{(j)};\bm{\phi})$

\State 4: \indent  Get $\{\bm{\theta}^{(j)}\}_{j=1}^{m}$ using $\bm{\theta}^{(j)} \leftarrow \mathcal{B}(\bm{\xi}^{(j)})$

\State 5: \indent Compute $\mathcal{L}$:
\begin{equation*}
\begin{split}
     \mathcal{L} &\leftarrow \frac{1}{m}\sum_{j=1}^{j=m} \Big[\log p(\mathbf{D}| \bm{\theta}^{(j)})+ \log p_{\bm{\theta}}(\bm{\theta}^{(j)})+ \mathcal{V}(\bm{\theta}^{(j)}, \bm{\xi}^{(j)}) - \log q_{\bm{\xi}}(\bm{\xi}^{(j)} ;\bm{\phi}) \Big] 
\end{split}
\end{equation*} 
 
\State 6: \indent Update $\bm{\phi}$ by the gradient: $\nabla_{\bm{\phi}}\mathcal{L}$ 
\State 7: \textbf{end while}
\end{algorithmic}
\end{algorithm}

In the following section, we will discuss modeling the conditional probability density $w_{s_{i}}(s_{i}|\theta_{i})$ and theoretical justifications of the TVI algorithm.

\subsection{Theoretical Justification}
\label{section:Theory}

We begin with Remark ~\ref{approx_error}, which provides a sufficient condition for the exact computation of a density function for $\bm{\theta}$ as the sum of the corresponding density function for $\bm{\xi}$ and the likelihood contribution $\mathcal{V}(\bm{\theta}, \bm{\xi})$ \citep{nielsen2020survae}.
\begin{Rem} 
\label{approx_error}
Consider $h_{\bm{\theta}}(\bm{\theta})$ and $h_{\bm{\xi}}(\bm{\xi})$ be some probability densities in the constrained and unconstrained space, respectively. If the support of the $h_{\bm{\xi}}(\bm{\xi})$ is a subset of $\mathcal{B}^{-1}(\bm{\Theta})$, then $\log h_{\bm{\theta}}(\bm{\theta}) =\log h_{\bm{\xi}}(\bm{\xi}) - \mathcal{V}(\bm{\theta}, \bm{\xi})$. 
\end{Rem}

Let $h_{\bm{\theta}}(\bm{\theta})$ and $h_{\bm{\xi}}(\bm{\xi})$ be $q_{\bm{\theta}}(\bm{\theta};\bm{\phi})$ and $q_{\bm{\xi}}(\bm{\xi};\bm{\phi})$, then the sufficient condition to have no approximation error, i.e., 
\vspace{-5pt}
\begin{equation*}
    \log q_{\bm{\theta}}(\bm{\theta};\bm{\phi}) =\log q_{\bm{\xi}}(\bm{\xi};\bm{\phi}) - \mathcal{V}(\bm{\theta}, \bm{\xi}),
\end{equation*} 
is to let the support of $q_{\bm{\xi}}(\bm{\xi};\bm{\phi})$ be a subset of $\mathcal{B}^{-1}(\bm{\Theta})$. To ensure that this condition is satisfied throughout optimization of $q_{\bm{\xi}}(\bm{\xi} ; \bm{\phi})$, some additional restriction is needed on the definition of surjection. 

If we set $h_{\bm{\theta}}(\bm{\theta})$ and $h_{\bm{\xi}}(\bm{\xi})$ in Remark~\ref{approx_error} to be $p_{\bm{\theta}}(\bm{\theta}|\mathbf{D})$ and $p_{\bm{\xi}}(\bm{\xi}|\mathbf{D})$, the support of $p_{\bm{\xi}}(\bm{\xi}|\mathbf{D})$ is a subset of $\mathcal{B}^{-1}(\bm{\Theta})$ by the definition of $\mathcal{B}$ in Section \ref{section:surjection} and hence we automatically have 
\begin{equation*}
    \log p_{\bm{\theta}}(\bm{\theta}|\mathbf{D}) =\log p_{\bm{\xi}}(\bm{\xi}|\mathbf{D}) - \mathcal{V}(\bm{\theta}, \bm{\xi}).
\end{equation*} 
However, some caution is still needed; the induced target density $p_{\bm{\xi}}(\bm{\xi} | \mathbf{D})$, given by the boundary surjection, should not cause additional challenges in optimization. For example, a poorly curated boundary surjection may result in extra mode or discontinuity in $p_{\bm{\xi}}(\bm{\xi} | \mathbf{D})$, which makes it difficult to optimize $\bm{\phi}$ in $q_{\bm{\xi}}(\bm{\xi};\bm{\phi})$.



To address these challenges, $w_{s_{i}}(s_{i}|\theta_{i})$ needs to be carefully designed. Our strategy is to introduce a boundary surjection radius, $r$. The main idea is to concentrate probability mass of $p_{\bm{\xi}}(\bm{\xi} | \mathbf{D})$ inside $\prod_{i=1}^{d} (a_{i}-r, b_{i}+r)$, which ensures that the support of $q_{\bm{\xi}}(\bm{\xi} ; \bm{\phi})$ belongs to $\mathcal{B}^{-1}(\bm{\Theta})$. Moreover, we impose a set of conditions, including smoothness and continuity conditions to $w_{s_{i}}(s_{i}|\theta_{i})$, which prevents having extra modes or discontinuities. For ease of exposition, we first consider the left side of the parameter boundary. We define $w_{s_{i}}(s_{i}=1|\theta_{i})= u_{i}(\theta_{i})$, and $w_{s_{i}}(s_{i}=0|\theta_{i})= 1 - u_{i}(\theta_{i})$ where function $u$ satisfies the following proposition. 

\begin{prop} 
\label{function_u}
Consider the $d$-dimensional hypercubes: $\mathcal{Q}_{r} (\bm{a}) = \prod_{i=1}^{d} [a_{i}-r, a_{i}+r]$, $\mathcal{Q}_{r-} (\bm{a}) = \prod_{i=1}^{d} [a_{i}-r, a_{i}]$, and $\mathcal{Q}_{r+} (\bm{a}) = \prod_{i=1}^{d} [a_{i}, a_{i}+r]$, which are obtained from the left side of the parameter boundary $\bm{a} = (a_{1},a_{2},...,a_{d})$ and for some radius $r \in (0, \min_{i}\{\frac{b_{i} - a_{i}}{2}\})$. Let $p_{\bm{\theta}}(\bm{\theta}| \mathbf{D})$ be the target density such that $0 < p_{\bm{\theta}}(\bm{\theta}| \mathbf{D}) \leq M < \infty$ and continuous for $\bm{\theta} \in \mathcal{Q}_{r+} (\bm{a})$. Let $q_{\bm{\xi}}(\bm{\xi} ; \bm{\phi})$ be the approximate density that satisfies $q_{\bm{\xi}}(\bm{\xi} ; \bm{\phi}) \in C^{m} \left(\mathcal{Q}_{r}(\bm{a}) \right)$; $C^{m} \left( \mathcal{S} \right)$ is the class of $m$ times differentiable functions over a set $\mathcal{S}$. If $u_{i}$ satisfies the following conditions for some $\epsilon > 0$ and $\forall i$,
\begin{enumerate}
\item $u_{i} \in {C}^{\infty}([a_{i}, \infty))$ such that $u_{i}:[a_{i}, \infty) \rightarrow [0.5,1)$ and monotone increasing, 
\item $u_{i}(a_{i}) = 0.5$,
\item $\forall R \in (r,\infty)$, $u_{i}(a_{i}+R) \geq 1 - \sqrt[\leftroot{-3}\uproot{3}d]{\epsilon/M}$,
\item $\forall \bm{c}= (c_{1},c_{2},...,c_{d}) \in \mathcal{Q}_{r+} (\bm{a})$, $\frac{\partial u_{i}(c_{i})}{\partial c_{i}}{(1-u_{i}(c_{i}))^{-1}} \geq \frac{\partial p_{\bm{\theta}}(\bm{c}| \mathbf{D})}{\partial c_{i}}{(p_{\bm{\theta}}(\bm{c}| \mathbf{D}))^{-1}}$,
\end{enumerate}
then $p_{\bm{\xi}}(\bm{\xi} | \mathbf{D}) \leq \epsilon$ for $\bm{\xi} \in \prod_{i=1}^{d} (-\infty, a_{i}-r) \cap \mathcal{B}^{-1}(\bm{\Theta})$,  $p_{\bm{\xi}}(\bm{\xi} | \mathbf{D})$ is continuous in $\mathcal{Q}_{r} (\bm{a})$, monotone increasing in $\mathcal{Q}_{r-} (\bm{a})$, and  $q_{\bm{\theta}}(\bm{\theta}; \bm{\phi}) \in C^{m} \left(\mathcal{Q}_{r+}(\bm{a}) \right)$.

\end{prop}

The proof of Proposition 1 is provided in the Supplementary Material. We now explain the implications of Proposition~\ref{function_u}. First, by constructing the right side of the parameter boundary similarly, i.e., defining $w(s_i=2|\theta_i)$ in a similar manner, the proposition forces the transformed target distribution to be light-tailed, with most probability mass concentrated in $\prod_{i=1}^{d}[a_{i}-r, b_{i}+r]$. Note that $\mathcal{B}^{-1}(\bm{\Theta})$ becomes $\prod_{i=1}^{d}(\frac{3a_{i}-b_{i}}{2}, \frac{3b_{i}-a_{i}}{2})$ for finite $a_i$ and $b_i$. If there is a one-sided parameter constraint (e.g., $\theta_i \in [a_i,\infty)$), $\mathcal{B}^{-1}(\bm{\Theta})=(-\infty,\infty)$. A light-tailed target distribution guides approximate distribution $q_{\bm{\xi}}(\bm{\xi} ;\bm{\phi})$ to concentrate its masses inside $\prod_{i=1}^{d}(a_{i}-r, b_{i}+r)$. This encourages the support of $q_{\bm{\xi}}(\bm{\xi} ;\bm{\phi})$ to be contained within $\mathcal{B}^{-1}(\bm{\Theta})$ during model training, ensuring that the conditions for $q_{\bm{\xi}}(\bm{\xi} ;\bm{\phi})$ in Proposition 1 is satisfied. Second, the boundary surjection does not cause additional difficulties in approximating $p_{\bm{\xi}}(\bm{\xi}|\mathbf{D})$. The conditions in Proposition 2 ensures that $p_{\bm{\xi}}(\bm{\xi}|\mathbf{D})$ is continuous, monotone increasing in $(a_{i}-r, a_{i})$, and monotone decreasing in $(b_{i}, b_{i}+r)$; therefore it does not introduce any extra modes. Inside the boundary, i.e., within $\prod_{i=1}^{d} (a_{i}+r, b_{i}-r)$, the transformed target $p_{\bm{\xi}}(\bm{\xi}|\mathbf{D})$ is identical to the original target $p_{\bm{\theta}}(\bm{\theta}|\mathbf{D})$, and thus there is no additional distortion due to the transformation within that range. Intuitively, the transformed target distribution in $\bm \Xi$ is a "smoothly unfolded distribution" of the target distribution in $\bm \Theta$ near the parameter boundary. Lastly, the boundary surjection does not limit the smoothness of the  $q_{\bm{\theta}}(\bm{\theta}; \bm{\phi})$ with Proposition \ref{function_u}. This condition ensures that $q_{\bm{\theta}}(\bm{\theta}; \bm{\phi})$ well approximates a target distribution, which typically has a smooth density function.   

As we pointed out, the success of boundary surjection depends on the choice of the function $u_{i}$ that should satisfy the conditions in Proposition~\ref{function_u}. In what follows, we introduce a family of functions helpful in constructing $u_{i}$. 

\begin{Rem}
\label{logistic_function}
Consider the location-scale family of the logistic function, $\varsigma_{i}(x;A, B) = \frac{1}{1+e^{-B(x-A)}}$. $\forall r \in (0, \frac{b_{i} - a_{i}}{2})$, we can find $A$ and $B$ such that $\varsigma_{i}(x;A, B)$ satisfies the first three conditions of Proposition \ref{function_u}. Also, the Condition~4 in Proposition~\ref{function_u} becomes
\begin{equation}
\label{monotone_logi_condition}
B \varsigma_{i}(c_{i};a_{i}, B) \geq \frac{\partial p_{\bm{\theta}}(\bm{c}| \mathbf{D})}{\partial c_{i}}{(p_{\bm{\theta}}(\bm{c}| \mathbf{D}))^{-1}}.
\end{equation}

\end{Rem}

Proof of Remark 2 is provided in the Supplementary Material. Remark \ref{logistic_function} demonstrates that a family of logistic functions can be used as the function $u$. We further note that as boundary surjection radius $r$ gets smaller, the left hand side in \eqref{monotone_logi_condition} gets bigger, while $\sup_{\bm{c} \in \mathcal{Q}_{r+}} \{\frac{\partial p_{\bm{\theta}}(\bm{c}| \mathbf{D})}{\partial c_{i}}{(p_{\bm{\theta}}(\bm{c}| \mathbf{D}))^{-1}}\}$ gets smaller. Therefore, Condition 4 in Proposition~\ref{function_u} holds for a sufficiently small $r$. 


We now revisit the example discussed in Section \ref{section:challenges}. As in \eqref{transformbound}, consider a parameter $\theta \in [0,1]$ with a boundary surjection function $\mathcal{B}$. The approximate density is now
\begin{align}
    q_{\theta}(\theta ; \bm{\phi}) &= q_{\xi}(\xi ; \bm{\phi}) \exp \big(-\mathcal{V}(\theta, \xi) \big) \label{transformbound_boundarysurjection}.
\end{align}
Compared to \eqref{transformbound}, $\exp \big(-\mathcal{V}(\xi, \theta) \big)$ does not become 0 and has a positive value at the parameter boundary of $\theta$. Hence, $q_{\theta}(\theta ; \bm{\phi})$ does not distort the estimated density, and therefore the constructed mapping does not have the boundary effect issue.

\subsection{Sequentially Annealed Posteriors}
\label{section:SAP}

Here, we propose a new temperature annealing scheme that modifies the standard NF framework by assigning a fixed sequence of annealed intermediate posterior distributions. We sequentially train each part of the model with the corresponding intermediate posterior in turn for a given pre-specified temperature schedule. To our knowledge, temperature annealing has not been utilized for variational inference with normalizing flows. 

Sequentially annealed posteriors (SAP) first sets the sequence of temperature $t^{(1)} > t^{(2)}, \dots, > t^{(K-1)} > t^{(K)} \equiv 1$. Then the target posterior density $p_{\bm{\theta}}(\bm{\theta} | \mathbf{D})$ with the $k$th temperature is 
\begin{equation}
    \log p_{\bm{\theta}^{(k)}}(\bm{\theta}^{(k)} | \mathbf{D}) = (1/t^{(k)})\log p(\mathbf{D}|\bm{\theta}^{(k)}) + \log p_{\bm{\theta}}(\bm{\theta}^{(k)}) + C(t^{(k)}),
    \label{target_distribution}
\end{equation}
where $C(t^{(k)})$ is a normalizing constant. The sequence of target distributions resembles that of annealed importance sampling (AIS) \citep{neal2001annealed}. The higher the temperature, the smaller the influence of the data information (likelihood) becomes, and hence the prior distribution density $p_{\bm{\theta}}(\bm{\theta})$ dominates. When the temperature reaches 1, the target distribution becomes the true posterior distribution with the density $p_{\bm{\theta}}(\bm{\theta}| \mathbf{D})$. The sequence of annealed target distributions is a natural extension of the NF model, which constructs a gradual change from $p_{\bm{\theta}}(\bm{\theta})$ to $p_{\bm{\theta}}(\bm{\theta}| \mathbf{D})$ (see Figure~\ref{Model_Outline}).




For each $t^{(k)}$, we introduce a block layer $f^{(k)}(\cdot ; \bm{\phi}^{(k)})$, which is a composite of multiple NF layers. Here, $\bm{\phi}^{(k)}$ is the variational parameter in the NFs belonging to the $k$th block. Using a sequential composite up to the $k$th block layer, we generate $\bm{\xi}^{(k)}$ from the latent distribution as 
\vspace{-10pt}
\begin{equation}
\begin{split}
    \bm{\xi}^{(k)} = (f^{(k)} \circ... f^{(2)} \circ f^{(1)})(\mathbf{z}_0;\bm{\phi}^{(k)}, ..., \bm{\phi}^{(2)}, \bm{\phi}^{(1)}), ~ \mathbf{z}_0 \sim N(\mathbf{0}_{d}, \mathbf{I}_{d}).
\end{split}
\label{block_distribution}
\end{equation}
We let $q_{\bm{\xi}^{(k)}}(\bm{\xi}^{(k)};\bm{\phi}^{(k)}, ...\bm{\phi}^{(1)})$ be the $k$th variational density of $\bm{\xi}^{(k)}$. By using the surjective function in \eqref{mappingdef}, we can transform $\bm{\theta}^{(k)}=\mathcal{B}(\bm{\xi}^{(k)})$. Then the variational objective for the $k$th block is given as follows:
\vspace{-5pt}{\small
\begin{equation}
\begin{split}
    \mathcal{L}^{(k)} = \mathbb{E}_{ q_{\bm{\theta}^{(k)}}(\bm{\theta}^{(k)};\bm{\phi}^{(k)}, ..., \bm{\phi}^{(1)})}[&(1/t^{(k)})\log p(\mathbf{D}|\bm{\theta}^{(k)}) + \log p_{\bm{\theta}}(\bm{\theta}^{(k)}) \\&- \log q_{\bm{\theta}^{(k)}}(\bm{\theta}^{(k)};\bm{\phi}^{(k)}, ..., \bm{\phi}^{(2)}, \bm{\phi}^{(1)}) + C(t^{(k)})]. 
\end{split}
\label{block_distribution_obj}
\vspace{-20pt}
\end{equation}}
Here we train each block sequentially and separately from the first to the last in order. For the $k$th block, we optimize $\bm{\phi}^{(k)}$ from \eqref{block_distribution_obj}, while fixing $\bm{\phi}^{(1)},\dots,\bm{\phi}^{(k-1)}$ obtained from the previous blocks. This allows us to drop the normalizing constant $C(t^{(k)})$ in \eqref{target_distribution} because the temperature $t^{(k)}$ is fixed within each block. This provides a substantial computational advantage compared to the adaptive temperature annealing methods \citep[e.g.][]{mandt2016variational}.


We initialize the parameters $\{\bm{\phi}^{(k)}\}$ to be zero, which makes the neural networks $\sigma_i$'s in \eqref{equation:NFoverall} as zero functions; thus, $f^{(k)}(\cdot ; \bm{\phi}^{(k)})$ becomes an identity function. As a result, training of the $k$th block can start from a well-trained function $f^{(k-1)} \circ... f^{(2)} \circ f^{(1)}$. By setting $t^{(k)}$ to be lower than $t^{(k-1)}$, the intermediate approximate density $q_{\bm{\theta}^{(k)}}(\bm{\theta}^{(k)};\bm{\phi}^{(k)}, ..., \bm{\phi}^{(2)}, \bm{\phi}^{(1)})$ reflects more information from the likelihood function $p(\mathbf{D}|\bm{\theta}^{(k)})$. 

We now explain the benefit of our annealing algorithm from the perspective of minimizing the reverse KL divergence in \eqref{VIeq0}. Without such annealing schemes, optimization based on reverse KL divergence can lead to local optima problems. This can happen especially when the approximate distribution covers only a part of the high-density region of the target distribution. Our approach alleviates this issue through a gradual learning strategy. When training the $k$th block, we initialize the $k$th block as $q_{\bm{\theta}^{(k)}}(\bm{\theta}^{(k)}; \bm{\phi}^{(k)}, \bm{\phi}^{(k-1)},..., \bm{\phi}^{(1)})=q_{\bm{\theta}^{(k-1)}}(\bm{\theta}^{(k)};\bm{\phi}^{(k-1)},..., \bm{\phi}^{(1)})$ because  $f^{(k)}(\cdot ; \bm{\phi}^{(k)})$ is initialized as an identity function. This reduces the distance between the approximate density $q_{\bm{\theta}^{(k)}}(\bm{\theta}^{(k)}; \bm{\phi}^{(k)}, \bm{\phi}^{(k-1)}..., \bm{\phi}^{(1)})$ and target density $p_{\bm{\theta}^{(k)}}(\bm{\theta}^{(k)}|\mathbf{D})$ at the beginning of training of the $k$th block. Compared to the standard NF model that initializes the approximate density $q_{\bm{\theta}}(\bm{\theta}; \bm{\phi})$ as $p_{\mathbf{z}_{0}}(\bm{\theta})$, our SAP approach is more effective in that it initializes from the approximate density resembling $p_{\bm{\theta}^{(k-1)}}(\bm{\theta}^{(k-1)}|\mathbf{D})$ that is much closer to $p_{\bm{\theta}^{(k)}}(\bm{\theta}^{(k)}|\mathbf{D})$.



\subsubsection{Weight Adjusted Fine Tuning}

Reverse KL divergence-based optimization is effective in capturing the mode of the target but underestimates the variance, thus losing accuracy in estimating the entire density \citep{regli2018alpha}. To see this, consider the definition of reverse KL divergence 
\begin{equation*}
\text{KL}( q_{\bm{\theta}}(\bm{\theta};\bm{\phi}) || \log p_{\bm{\theta}}(\bm{\theta}|\mathbf{D}))
 = \int_{\bm{\Theta}} q_{\bm{\theta}}(\bm{\theta};\bm{\phi}) \left[\log q_{\bm{\theta}}(\bm{\theta};\bm{\phi}) - \log p_{\bm{\theta}}(\bm{\theta}|\mathbf{D})\right] d(\bm{\theta}).
 \end{equation*}
 The log difference $\log q_{\bm{\theta}}(\bm{\theta};\bm{\phi}) - \log p_{\bm{\theta}}(\bm{\theta}|\mathbf{D})$ will be ignored outside of the support $q_{\bm{\theta}}(\bm{\theta};\bm{\phi})$. This often creates some local minima in terms of reverse KL divergence, where $q_{\bm{\theta}}(\bm{\theta};\bm{\phi})$ minimizes the difference $\log q_{\bm{\theta}}(\bm{\theta};\bm{\phi}) - \log p_{\bm{\theta}}(\bm{\theta}|\mathbf{D})$ only in a fraction of the support of $p_{\bm{\theta}}(\bm{\theta}|\mathbf{D})$. Learning algorithms based on gradient descent are often trapped in such local minima.

One possible way to mitigate this issue is to apply forward KL divergence, but it cannot be implemented in a na\"{i}ve manner as it requires samples from the true target density. Importance sampling approximations are often used to circumvent such issue \citep{li2016renyi, bornschein2014reweighted, jerfel2021variational}. 

Inspired by these methods, we adjust our reverse KL divergence as follows. We first define a  function $w: \Theta \xrightarrow[]{} \mathbb{R}$ , which is bounded by the lower $\alpha_1$ and upper bound $\alpha_2$ with $0\leq \alpha_1 \leq 1 \leq \alpha_2 < \infty$. $w$ is a monotonically increasing function of the log differences ${\varrho}(\bm{\theta}) =  \log q_{\bm{\theta}}(\bm{\theta};\bm{\phi}) - \log p_{\bm{\theta}}(\bm{\theta}|\mathbf{D})$, with $w(\bm{\theta})=\alpha_1=\alpha_2=1$ if  $\varrho(\bm{\theta}) = 0$. With the defined $w(\bm{\theta})$, the adjusted reverse KL divergence variational objective is defined as
\vspace{-8pt}
\begin{equation}
\begin{split}
    \mathcal{L}_{w} &= \mathbb{E}_{ q_{\bm{\theta}}(\bm{\theta};\bm{\phi})} \left[  w(\bm{\theta}) (\log p(\mathbf{D}| \bm{\theta}) + \log p_{\bm{\theta}}(\bm{\theta}) - \log q_{\bm{\theta}}(\bm{\theta};\bm{\phi})) \right] \\ &\approx \frac{1}{m}\sum_{j=1}^{m} w(\bm{\theta}^{(j)}) \Big[\log p(\mathbf{D}| \bm{\theta}^{(j)})+ \log p_{\bm{\theta}}(\bm{\theta}^{(j)}) - \log q_{\bm{\theta}}(\bm{\theta}^{(j)} ;\bm{\phi}) \Big].
\end{split}
\label{general divergence}
\end{equation}
which can be viewed as an adjusted KL divergence 
\begin{equation*}
\text{KL}_{w}( q_{\bm{\theta}}(\bm{\theta};\bm{\phi}) || \log p_{\bm{\theta}}(\bm{\theta}|\mathbf{D})) = \int_{\bm{\Theta}}  w(\bm{\theta})  q_{\bm{\theta}}(\bm{\theta};\bm{\phi}) \left[\log q_{\bm{\theta}}(\bm{\theta};\bm{\phi}) - \log p_{\bm{\theta}}(\bm{\theta}|\mathbf{D}))\right] d(\bm{\theta}).
\end{equation*}

To illustrate how this algorithm can mitigate the aforementioned problem of the reverse KL divergence, we first define $A_{+} = \{\bm{\theta}| \varrho(\bm{\theta}) > 0\}$ and $A_{-} = \{\bm{\theta}| \varrho(\bm{\theta}) < 0\}$, i.e., $A_{+}$ is the set where the approximate distribution overestimates the target density, and $A_{-}$ is where it underestimates the target density. The lower bound of the adjusted KL divergence can be written as follows:
\begin{equation}
    \begin{split}
    \text{KL}_{w}( q_{\bm{\theta}}(\bm{\theta};\bm{\phi}) || \log p_{\bm{\theta}}(\bm{\theta}|\mathbf{D})) 
    =& \int_{A_{-}}  w(\bm{\theta})  q_{\bm{\theta}}(\bm{\theta};\bm{\phi}) \left[\log q_{\bm{\theta}}(\bm{\theta};\bm{\phi}) - \log p_{\bm{\theta}}(\bm{\theta}|\mathbf{D}))\right] d(\bm{\theta}) 
    \\&+ \int_{A_{+}}   w(\bm{\theta})  q_{\bm{\theta}}(\bm{\theta};\bm{\phi}) \left[\log q_{\bm{\theta}}(\bm{\theta};\bm{\phi}) - \log p_{\bm{\theta}}(\bm{\theta}|\mathbf{D}))\right] d(\bm{\theta}) \\
    \geq &~
        \alpha_2   \int_{A_{-}}  q_{\bm{\theta}}(\bm{\theta};\bm{\phi}) \left[\log q_{\bm{\theta}}(\bm{\theta};\bm{\phi}) - \log p_{\bm{\theta}}(\bm{\theta}|\mathbf{D}))\right] d(\bm{\theta}) \\ &+ \alpha_1   \int_{A_{+}}  q_{\bm{\theta}}(\bm{\theta};\bm{\phi}) \left[\log q_{\bm{\theta}}(\bm{\theta};\bm{\phi}) - \log p_{\bm{\theta}}(\bm{\theta}|\mathbf{D}))\right] d(\bm{\theta})\\
    =& (\alpha_2 - \alpha_1)\int_{A_{-}}  q_{\bm{\theta}}(\bm{\theta};\bm{\phi}) \left[\log q_{\bm{\theta}}(\bm{\theta};\bm{\phi}) - \log p_{\bm{\theta}}(\bm{\theta}|\mathbf{D}))\right] d(\bm{\theta}) \\
    &+ \alpha_1 \underbrace{ \int_{\bm{\Theta}}  q_{\bm{\theta}}(\bm{\theta};\bm{\phi}) \left[\log q_{\bm{\theta}}(\bm{\theta};\bm{\phi}) - \log p_{\bm{\theta}}(\bm{\theta}|\mathbf{D}))\right] d(\bm{\theta}) \label{WRKL_final}}_{\text{KL}(q||p)}
    \end{split}
\end{equation}
During optimization, minimizing the divergence defined in \eqref{general divergence} encourages the approximate distribution $q_{\bm{\theta}}(\bm{\theta};\bm{\phi})$ to shift the probability mass of $A_{+}$ to $A_{-}$ because  we are giving extra weight $\alpha_2 - \alpha_1>0$ to the term $\int_{A_{-}}  q_{\bm{\theta}}(\bm{\theta};\bm{\phi}) \left[\log q_{\bm{\theta}}(\bm{\theta};\bm{\phi}) - \log p_{\bm{\theta}}(\bm{\theta}|\mathbf{D}))\right] d(\bm{\theta})$. If the algorithm eventually reaches $q_{\bm{\theta}}(\bm{\theta};\bm{\phi}) = p_{\bm{\theta}}(\bm{\theta}|\mathbf{D}))$  for all $\bm{\theta}$, $w(\bm{\theta})=\alpha_1=\alpha_2=1$ and we have 
\begin{equation*}
\text{KL}_{w}( q_{\bm{\theta}}(\bm{\theta};\bm{\phi}) || \log p_{\bm{\theta}}(\bm{\theta}|\mathbf{D})) =\text{KL}( q_{\bm{\theta}}(\bm{\theta};\bm{\phi}) || \log p_{\bm{\theta}}(\bm{\theta}|\mathbf{D})), 
\end{equation*}
eliminating the extra weight term $(\alpha_2 - \alpha_1)\int_{A_{-}}  q_{\bm{\theta}}(\bm{\theta};\bm{\phi}) \left[\log q_{\bm{\theta}}(\bm{\theta};\bm{\phi}) - \log p_{\bm{\theta}}(\bm{\theta}|\mathbf{D}))\right] d(\bm{\theta})$. In our implementation, we define $w(\bm{\theta}^{(j)})= \frac{p_{\bm{\theta}}(\bm{\theta}^{(j)}|\mathbf{D})}{q_{\bm{\theta}}(\bm{\theta}^{(j)};\bm{\widehat{\phi})}}$ for the $j$th sample in \eqref{WRKL_final}. Note that $w(\bm{\theta}^{(j)})$ is not considered as a function of $\bm \phi$ when deriving the gradients, as it is treated as the importance weight given by the estimated value $\bm{\widehat{\phi}}$. In our optimization algorithm, we set $\bm{\widehat{\phi}}$ to be an estimated value in the previous iteration.

In practice, this method is still not applicable when $q_{\bm{\theta}}(\bm{\theta};\bm{\phi})$ is far away from  $p_{\bm{\theta}}(\bm{\theta} | \mathbf D)$; the variance of $w(\bm{\theta})$ goes to infinity, and self normalization will likely to truncate the smaller values to zero especially in high dimension. To stabilize these weights, we use Pareto smoothed importance sampling (PSIS) for weight normalization \citep{vehtari2015pareto}. This method calculates truncated normalized importance weight $\Tilde{w}(\bm{\theta})$ which is asymptotically unbiased and consistent with finite weight variance. Moreover, $\Tilde{w}(\bm{\theta})$ satisfies all the conditions for $w(\bm{\theta})$ described above as it is an increasing function of  ${\varrho}(\bm{\theta})$ and becomes 1 when ${\varrho}(\bm{\theta})=0$. 

One drawback is the use of PSIS increases the computational cost because the important sampling size has to be sufficiently large to ensure unbiasedness. To circumvent this issue, we apply forward KL with PSIS only after the convergence of the reverse KL divergence objective. This strategy can reduce the computational cost as well as stabilize importance sampling weights. We propose the Annealed and Transformed Variational Inference (ATVI) algorithm, which is summarized in Algorithm~2.

\begin{algorithm}[ht]
\caption{Annealed and transformed variational inference (ATVI) algorithm }\label{alg}
\begin{algorithmic}
\small
\State 1: Initialize $\bm{\phi}$ and set the 
 temperature ladder $\{t^{(k)}\}_{k=1}^{K}$.
\State 2: \textbf{for} $k = 1$ to $K$ \textbf{do}
\State 3: \indent \textbf{while} \textit{$\mathcal{L}^{(k)}$ not converged} \textbf{do}
\State 4: \indent \indent Sample $\{\mathbf{z}_{0}^{(j)}\}_{j=1}^{m} \sim p_{\mathbf{z}_{0}}(\mathbf{z}_{0})$ and get $\{\bm{\xi}^{(k),(j)}\}_{j=1}^{m}$ using \eqref{block_distribution}
\State 5: \indent \indent Get $\{\bm{\theta}^{(k),(j)}\}_{j=1}^{m}$ using $\bm{\theta}^{(k),(j)} \leftarrow \mathcal{B}(\bm{\xi}^{(k),(j)})$
\State 6: \indent \indent Calculate $\mathcal{L}^{(k)}$:
\begin{equation*}
\begin{split}
      \mathcal{L}^{(k)} \leftarrow &\frac{1}{m}\sum_{j=1}^{m} \Big[\frac{1}{t^{(k)}}\log p(\mathbf{D}| \bm{\theta}^{(k),(j)}) \\
     &+ \log p_{\bm{\theta}}(\bm{\theta}^{(k),(j)}) + \mathcal{V}(\bm{\theta}^{(k),(j)}, \bm{\xi}^{(k),(j)}) - \log q_{\bm{\xi}}(\bm{\xi}^{(k),(j)} ;\bm{\phi}) \Big] 
\end{split}
\end{equation*} 
\State 7: \indent \textbf{end while}
\State 8: \textbf{end for}

\State 9: \textbf{while} \textit{$\mathcal{L}_{w}$ not converged} \textbf{do}: (\textbf{weight adjusted fine-tuning})

\State 10: \indent Calculate truncated importance weight $\tilde{w}(\bm{\theta}^{(j)})$ using PSIS
\State 11: \indent Calculate $\mathcal{L}_{w}^{(k)}$:
\begin{equation*}
\begin{split}
 \mathcal{L}_{w}^{(k)}  \leftarrow &\frac{1}{m}\sum_{j=1}^{m}\tilde{w}(\bm{\theta}^{(j)}) \Big[\frac{1}{t^{(k)}}\log p(\mathbf{D}| \bm{\theta}^{(k),(j)})+ \log p_{\bm{\theta}}(\bm{\theta}^{(k),(j)}) \\
     & + \mathcal{V}(\bm{\theta}^{(k),(j)}, \bm{\xi}^{(k),(j)}) - \log q_{\bm{\xi}}(\bm{\xi}^{(k),(j)} ;\bm{\phi}) \Big] 
\end{split}
\end{equation*} 
\State 12: \indent Update $\bm{\phi}^{(k)}$ by the gradient: $\nabla_{\bm{\phi}^{(k)}}\mathcal{L}^{(k)}$ 
\State 13: \textbf{end while}
\end{algorithmic}
\end{algorithm}

\section{Applications}
\label{Applications}

In this section, we apply our approach to various calibration examples. To illustrate the performance of our method, we compare ATVI with several baseline Bayes approaches. As a benchmark representing the conventional MCMC approach, we use the adaptive Metropolis sampler proposed by \cite{vihola2012robust}. The method seeks an adaptive matrix factor that utilizes the shape of the target parameter that achieves a given acceptance rate. The method is known to be robust and served as an inference engine in various papers \citep[see, e.g.,][]{park2017ensemble, akiyama2022first}. We also include a na\"{i}ve version of our NF-based approach  (without annealing and boundary surjection), denoted as VI (NF) henceforth, and Gaussian copula method with a G\&H transformation \citep{smith2020high}, denoted as VI (copula), as benchmark models in our comparison. We implement all algorithms regarding VI in {\tt{PyTorch}} and implement the adaptive MCMC method in {\tt{R}}.


Except for the MCMC algorithm used in Section 5.2, the computation times are calculated with a machine with AMD Ryzen 5 5600X 6-Core Processor 3.70 GHz and NVIDIA GeForce RTX 5000. For the MCMC algorithm in Section 5.2, the code is implemented on AMD Ryzen 9 3900X 12-Core Processor 3.79 GHz to accelerate computation. For VI (NF) and ATVI methods, the learning rates are selected among the values of $\{0.8, 1, 2, 3, 6, 8, 10, 100\}\times 10^{-3}$, to balance the speed of convergence and estimation accuracy, which is a common practice in the optimization literature.

Regarding the choice of the temperature ladder, we found that a temperature schedule with only two steps $(t^{(1)},t^{(2)})$ is enough in all three examples. Here $t^{(2)}$ is always set to 1 by definition. For $t^{(1)}$, to avoid overly complicated tuning that might lead to a high computational burden, we applied a simple rule of thumb, vaguely inspired by the commonly used geometric spacing: we tested four values, $1$, $1.7 (\approx\sqrt{3})$, $3$, and $5 (\approx 3\sqrt{3})$, and selected the one that yielded the best result in terms of data coverage.


\subsection{SIR Model Example}


Susceptible-infected-recovered (SIR) models \citep{kermack1927contribution, dietz1967epidemics} are a popular class of compartmental models for studying infectious disease dynamics. Hosts within a population are categorized as susceptible ($S$), infected ($I$), and recovered/removed ($R$), which are mutually exclusive groups. If hosts are exposed to diseases but are not infected yet, they are susceptible. They are infected if the pathogen has already spread to them. If their immune system successfully removes the pathogen, they are classified as recovered (or "removed" if they die). Let $S, I, R$ denote the number of susceptible, infected, and recovered hosts within a population size $N$ ($S+I +R = N$). Then
the basic (deterministic) SIR model has the following ODE:
\begin{equation}
\small
\begin{split}
& \frac{\partial S}{\partial t} = -\frac{\beta I}{N}S \\
& \frac{\partial I}{\partial t} = \frac{\beta I}{N}S - \gamma I \\
& \frac{\partial R}{\partial t} = \gamma I.
\label{sir}
\end{split}
\end{equation}
The dynamics resemble the intuitive movement of the epidemic, as model parameters control the population transitions between each compartment. 

Here, we calibrate the basic SIR model in \eqref{sir} using the common-cold prevalence data spanning 21 days on Tristan da Cunha in October 1967, as obtained by \cite{shibli1971common}. We assume that both the infected and recovered patients are observed, thus we denote $\mathbf{D}=\lbrace D_{i,t}; t \in (1,\cdots,21), i \in (1, 2) \rbrace$ as the observed data, where $D_{1,t}$ represents the number of infected individuals and $D_{2,t}$ represents the number of recovered individuals. The likelihood for $D_{1,t}$ and $D_{2,t}$ is defined based on the Poisson distribution with means $I_{t}$ and $R_{t}$ respectively.


The parameters that needs to be estimated are $\beta$ and $\gamma$ in \eqref{sir} as well as $S_0$, the initial condtion for $S_0$ (the initial conditions for $I$ and $R$ are set to be 0).
We use uniform priors for all three parameters over a plausible range. Since the total infected population is over 38 people, we set the lower bound for $S_0$ to be 38. We provide details about the priors in the Supplementary Material (Table S1).  For VI (NF) and ATVI, we use 20 rational quadratic flows (10 for $t^{(1)}=3$ and 10 for $t^{(2)}=1$) with about 350 gradient updates. We set the learning rate as $3 \times 10^{-3}$ for both approaches, which led to the convergene of the loss function in our example. 


\begin{table}[h]
\centering
\small
\begin{tabular}{ccccc}
\hline
& ATVI & VI (NF) & VI (copula) & MCMC \\
\hline
$\beta$ & 0.87 & 0.87 &0.86 & 0.89\\
95\% HPD & (0.83, 0.95) & (0.80, 0.96) & (0.81, 0.96) &(0.82, 0.96) \\
\hline
$\gamma$ & 0.30 & 0.31 & 0.30 & 0.29\\
95\% HPD & (0.27, 0.35) & (0.27, 0.37) & (0.26, 0.39) & (0.24, 0.34) \\
\hline
$S_{0}$  & 39.47 & 42.62 & 42.57 & 39.37\\
95\% HPD & (37.03, 41.61) & (40.16, 46.71) & (39.32, 48.25) & (37.00, 43.53)\\
\hline
Comp. Time (mins) & 1.7 & 1.2 & 1.2 & 1.2\\
Data Coverage (\%) &97.6& 100&97.6  & 97.6 \\
Data AIL &254&267.5&264.0&254.5\\ 
Data MSPE &224.0&241.0&241.6&228.7\\
\hline
\end{tabular}
\caption{Inference results for the common cold dataset. For all parameters, mode of estimates, 95\% HPD interval (in parenthesis), and computing time (minutes) are reported. For the forward simulation results, empirical coverage (Data Coverage) the average length of the interval estimates (AIL), and mean squared prediction errors (MSPE) are reported.}
\label{table:tablesir_result}
\end{table}

\begin{figure}[h]
\begin{center}
    \includegraphics[ scale = 0.4]{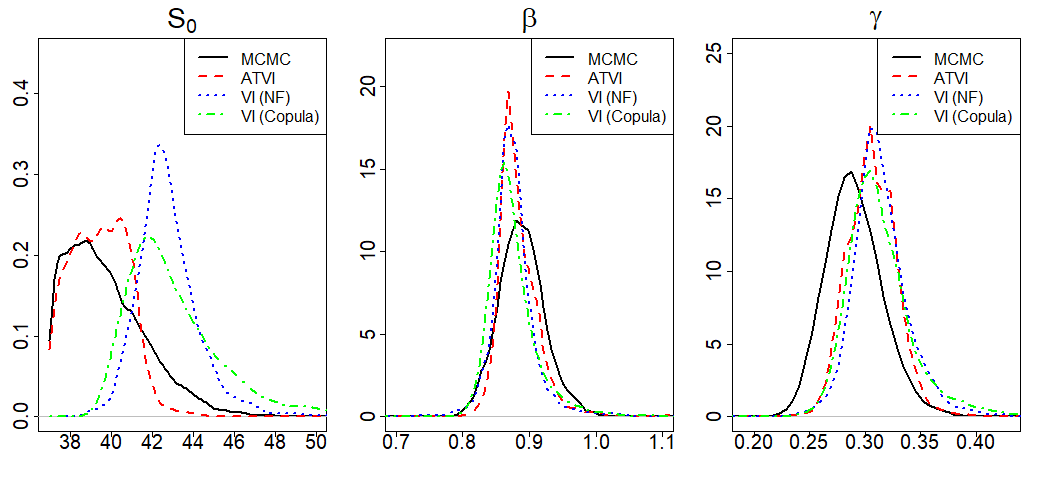}
\end{center}
\caption[]{The inference results of the common cold dataset. For all figures, black line indicates the density of the MCMC method, red line for the ATVI method, blue line for the VI with normalizing flow method and the green for the VI with copula method. }
\label{Param_SIR_figure}
\end{figure}

The parameter estimation results in Table \ref{table:tablesir_result} and Figure \ref{Param_SIR_figure} show that the estimation results for $S_0$ by VI (NF) and VI (copula) are clearly affected by the boundary effects in estimating $S_0$, but the result by ATVI does not show such an issue, yielding a result quite close to the adaptive MCMC. The forward simulation results based on the estimated parameters, illustrated in Figure S1 in the Supplementary Material, show that all four compared methods lead to visually similar results, but the average interval length (AIL) and the mean squared prediction error (MSPE) in Table \ref{table:tablesir_result} indicate that ATVI yields the closest forward simulation trajectories of the infected and recovered groups to the adaptive MCMC.

\vspace{-10pt}
\subsection{MSIR Model Example}
\label{section:MSIR} 
Rotavirus is the primary reason for diarrheal disease among children and is one of the dominant causes of death \citep{parashar2006rotavirus}. However, studying rotavirus dynamics is challenging due to the lack of diagnostic capacity and under-reporting issues; therefore, noisy data are often observed, resulting in difficulties in parameter estimation. Recently, \cite{park2017ensemble} analyzed the rotavirus disease for children in Niger with several variants of compartment models. Here, we examine the MSIR model studied in \cite{park2017ensemble}.

The model considers age-structured transmission rates in six groups: 0-1 month, 2-3 months, 4-5 months, 6-11 months, 12-23 months, and 24-59 months. A time-varying transmission rate is $\beta_{i}(t) = \beta_{0i}\Big(1 + w \cos\Big(\frac{2\pi t - 52 \phi}{52}\Big)\Big)$, where $\beta_{0i}$ is the baseline rate parameter for age class $i$, $\phi$ is the offset of the seasonal variation, and $w$ is an amplitude parameter. Then the force of infection, the rate of susceptibles acquiring rotavirus disease, can be computed as $\lambda_{i}=\sum_{j=1}^{6} \beta_{j}(t) C_{ij} \frac{(I^{(s)} +0.5I^{(m)})}{N_{j}}$, where $C_{ij}$ is the frequency of contact between the age classes $i$ and $j$, and $N_j$ is the number of population in the age class $j$. $I^{(s)}$ and $I^{(m)}$ represent groups infected with severe and mild rotavirus, respectively. Then the model has the following ODE:
\vspace{-8pt}{\small
\begin{align}
    & \frac{\partial M_{i}}{\partial t} = \alpha_{i-1}M_{i-1} - \alpha_{i}M_{i} + \mu N - \frac{1}{13} M_{i} \label{eqm1}\\ 
    & \frac{\partial S_{i}}{\partial t} = \alpha_{i-1}S_{i-1} - \alpha_{i}S_{i} + \frac{1}{13} M_{i} - \lambda_{i} S_{i} + \frac{1}{52} R_{i} \label{eqm2}\\ 
    & \frac{\partial I_{i}^{(s)}}{\partial t} = \alpha_{i-1}I_{i-1}^{(s)} - \alpha_{i}I_{i}^{(s)} + \frac{6}{25}\lambda_{i} S_{i} -  I_{i}^{(s)} \label{eqm3}\\
    & \frac{\partial I_{i}^{(m)}}{\partial t} = \alpha_{i-1}I_{i-1}^{(m)} - \alpha_{i}I_{i}^{(m)} + \frac{19}{25}\lambda_{i} S_{i} - 2 I_{i}^{(m)}. \label{eqm4}  \vspace{-8pt}  
\end{align}}
Here, \eqref{eqm1} describes the maternal immunity ($M_i$), which is assumed to be 13 weeks. Newborns are added to $M_i$ with the birth rate $\mu$. After a period of maternal immunity, the individual child becomes susceptible ($S_i$) as in  \eqref{eqm2}. Then susceptible individuals can be infected with mild ($I_{i}^{(m)}$) or severe ($I_{i}^{(s)}$) rotavirus as described in \eqref{eqm3} and \eqref{eqm4}, respectively. Individuals can leave each state for a mean period of a week (severe) or half a week (mild). If infected children are successfully immunized, they are classified as recovered ($R_i$); even after recovery, individuals may reenter susceptible with a rate of $1/52$ (i.e., the mean period of immunity following infection is assumed to be 52 weeks). $\bm{\alpha}=\lbrace 1/8, 1/8, 1/8, 1/24, 1/48, 1/144 \rbrace$ represents the movement rates between age classes. Further model details, including $C_{ij}$ and $\mu$ are provided in Section S2.1 of the Supplementary Material. Let $\mathbf{D}=\lbrace D_{i,t}; t \in (1,\cdots,118), i \in (1,\cdots,6) \rbrace$ be the observed data, where $D_{i,t}$ is the number of reported rotavirus cases in age class $i$ during week $t$. Then $D_{i,t}$ is modeled via negative binomial distribution with mean $\eta_i(t) = \frac{6}{25}\rho \lambda_{i,t}S_{i,t}$ and dispersion rate $\nu$. Here, $\eta_i(t)$ is called a disease burden, and $\rho$ is a reporting rate. 

We use synthetic observations that have very similar statistical characteristics to the real observations to avoid data protection issues. We simulated an epidemic from the MSIR model for given true parameters $w=0.43, \phi=7.35, \rho=0.027, \nu=0.9, \beta_{01}= 1.305,\beta_{02}= 3.118, \beta_{03}=0.032, \beta_{04}=3.241, \beta_{05}=0.314$, and $\beta_{06}=0.094$; to consider realistic settings, we used the estimated parameter values in \cite{park2017ensemble}. We used uniform priors of model parameters over a plausible range as in \cite{park2017ensemble}. Details are provided in the Supplementary Material (Table S2).

For the NF approaches, ATVI, we use 20 layers (10 for $t^{(1)}=3$ and 10 for $t^{(2)}=1$) of the flow model with the learning rate of $3 \times 10^{-3}$. Similarly, VI (NF) use 20 layers. For VI (copula), the learning rate is set to be $1 \times 10^{-1}$. 
We run the adaptive MCMC algorithm for 50,000 iterations, which can generate effective sample sizes between 300 and 1200, depending on the parameters. 


\begin{table}[h]
\centering
\small
\begin{tabular}{ccccc}
\hline
& ATVI & VI (NF)& VI (copula) & MCMC \\
\hline
$w$ & 0.450 (0.002) & 0.457 (0.005) & 0.322 (0.043)& 0.467 (0.004)\\
Coverage (\%) & 82 & 40 &86& 94\\
\hline
$\phi$ & 7.386 (0.186) & 6.621 (1.900) &7.47 (0.025)& 7.416 (0.012)\\
Coverage (\%) & 90 & 60 &100& 88\\
\hline
$\rho$  & 0.026 ($0.870^{*}$)  & 0.024 ($3.132^{*}$) &0.022 ($1.711^{*}$)& 0.027 ($2.667 ^{*}$) \\
Coverage (\%) & 84 & 36 &60& 96\\
\hline
$\nu$ & 0.868 (0.007) & 0.809 (0.017) &0.870(0.026)& 0.902 (0.005) \\
Coverage (\%) & 86 & 46 &90& 100 \\
\hline
$\beta_{01}$ & 1.411 (0.260) & 1.440 (0.192) &1.093(0.597)& 1.785 (1.300) \\
Coverage (\%) & 94 & 92 &100& 80\\
\hline
$\beta_{02}$ & 2.844 (0.280) & 2.826 (0.226) &3.381 (0.348)& 3.258 (0.470) \\
Coverage (\%) & 90 & 92 &100& 98\\
\hline
$\beta_{03}$ & 0.033 (0.006) & 0.091 (0.093) &0.050 (0.003)& 0.397 (1.224) \\
Coverage (\%) & 96 & 62 &96& 94\\
\hline
$\beta_{04}$ & 2.903 (0.342) & 2.972 (0.219) &3.479 (0.505) & 2.997 (0.578) \\
Coverage (\%) & 80 & 94 &100& 92\\
\hline
$\beta_{05}$ & 0.419 (0.355) & 0.657 (0.435) &0.4957 (1.136)& 0.335 (0.013) \\
Coverage (\%) & 96 & 48 &100& 98\\
\hline
$\beta_{06}$ & 0.220 (0.569) & 0.479 (0.681) &0.843 (3.939)& 0.101 (0.001) \\
Coverage (\%) & 94 & 36 &60& 96\\
\hline
Comp. Time (hours) & 1.9 & 1.1 &1.6& 15.2\\
Data Coverage (\%) & 96.6 & 96.5 &96.2& 94.3\\
Data AIL &84.72& 88.38&91.72&84.78\\
Data MSPE &1429.16& 1803.74&1689.69&1523.71\\
Burden Coverage (\%) & 86.0 & 59.0 &62.0& 93.8\\ 
Burden AIL &24.07& 17.37&25.58&25.14\\
Burden MSPE  &112.11& 221.75&260.142&111.74\\
\hline
\end{tabular}
\caption{Inference results for the simulated rotavirus example from 50 repeated simulations. For all parameters, posterior mode of estimates, MSE for the estimates (in paraethesis), empirical coverage for parameters are reported. The average computing time and forward simulation results are also reported. Data Coverage, AIL, MSPE are results for the severe cases summed over all age groups. Burden coverage, AIL and MSPE are results for the diease burden estimates, summed over the six age groups.\\
* These are MSE values multiplied by $10^{5}$.}
\label{table:tablemsir_result}
\end{table}


We repeated the simulation 50 times to evaluate the compared methods, and the results are summarized in Table~\ref{table:tablemsir_result}. The results show that the empirical coverage of VI (NF) is far below the nominal coverage for all four non-age-specific parameters ($w$, $\phi$, $\rho$, and $\nu$), as well as three age-specific parameters ($\beta_{03}$, $\beta_{05}$, and $\beta_{06}$), indicating that the model suffers from a local minima issue. VI (copula) shows better results than VI (NF), but it notably underestimates the amplitude parameter ($w$) and has under-coverage issues for $\rho$ and $\beta_{06}$. On the other hand, ATVI shows more stable results across all parameters, more similar to the adaptive MCMC. Figure S2 illustrates estimated parameter densities from one of the repeated simulations, which demonstrates an example iteration where ATVI more reliably estimates all the parameters than the other VI methods.



\begin{figure}[ht]
\begin{center}
    \begin{tabular}{c c}
        \textbf{ATVI} & \textbf{VI (NF)} \\
        \includegraphics[width=0.36\columnwidth]{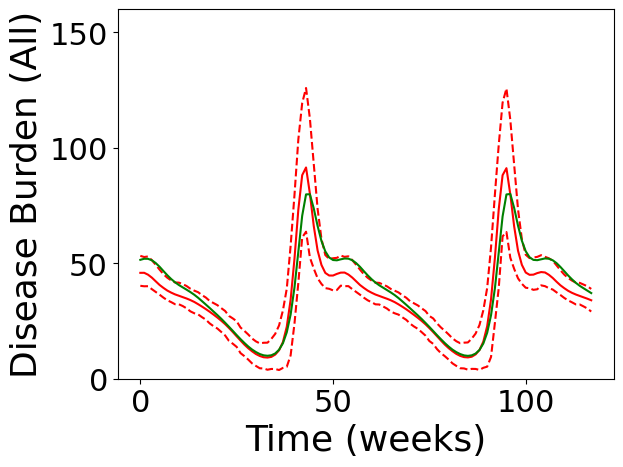} & \includegraphics[width=0.36\columnwidth]{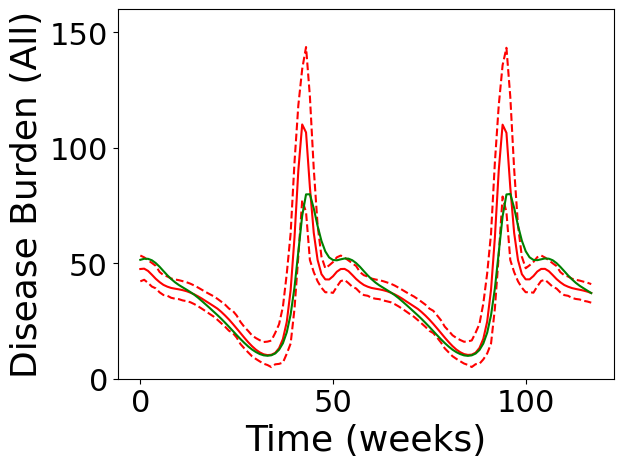} \\
        \textbf{VI (copula)}& \textbf{MCMC} \\ 
        \includegraphics[width=0.36\columnwidth]{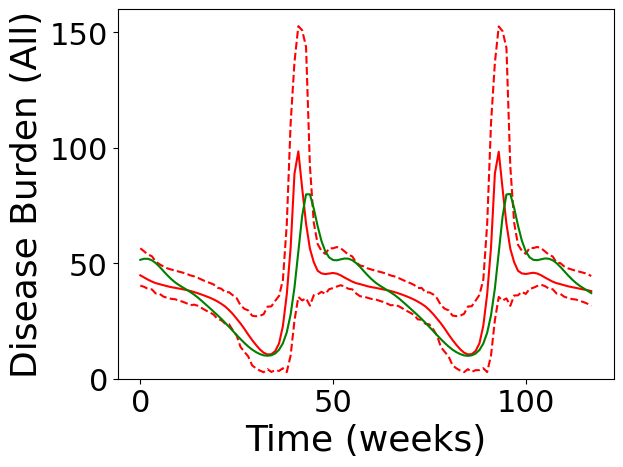}
         & \includegraphics[width=0.36\columnwidth]{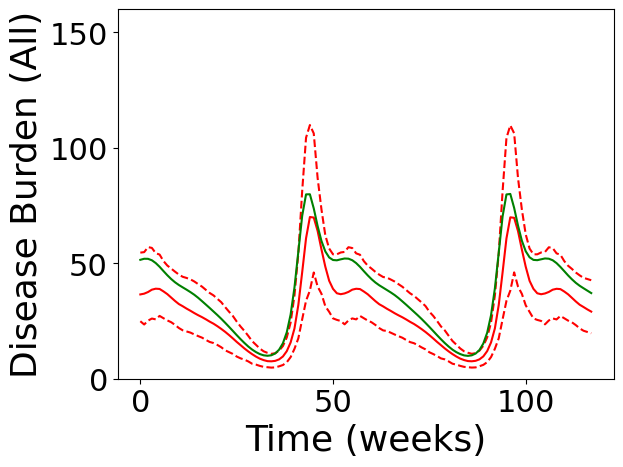} \\ 
    \end{tabular}
\end{center}
\caption[]{The burden estimate (red lines) with the true disease burden (green lines). For all figures, solid and dashed lines indicate posterior mean and corresponding 95\% HPD interval, respectively. }
\label{Burden_MSIR_figure}
\end{figure}

As in the previous example, we also evaluate the prediction performance by running forward simulations based on the parameter estimates. We first obtain a posterior sample of the total disease burden $\eta(t)=\sum_{i=1}^6 \eta_i(t)$ based on the samples for the parameters. Additionally, we generate simulated severe cases from the negative binomial distribution with mean $\eta_{i}(t)$ and dispersion $\nu$. From this, we can obtain the posterior means and the 95\% HPD intervals for both the disease burden and the generated observations. Then we compute the empirical coverage of the resulting intervals for their corresponding target quantities as well as the average interval lengths. The results in Table~\ref{table:tablemsir_result} show that while all methods can achieve the nominal coverage for the observed data, only ATVI can achieve the MSPE and AIL comparable to the adaptive MCMC. For the burden estimate, only MCMC can achieve the nominal coverage for the burden estimate (93.8\%). However, ATVI leads to a much higher coverage (84.9\%) than the other two VI methods (59.0\% for NF and 62\% for copula), and its MPSE is comparable to MCMC. Figure S3 in the Supplementary Material for observed severe cases and Figure \ref{Burden_MSIR_figure} for the disease burden from an example iteration demonstrate a case where ATVI produces the closest forward simulation results to those from the adaptive MCMC.

\vspace{-7pt}
\subsection{WRF-Hydro Model Example}
\label{section:WRF} 

In this section, we turn our attention to the problem of calibrating a model used in a different field from the previous examples. The Weather Research and Forecasting Model Hydrological modeling system (WRF-Hydro) \cite{gochis2018wrf} is a geophysical model developed by the National Center of Atmospheric Research (NCAR). The model provides a unified approach to simulating weather, land surface, and high-resolution hydrologic processes. Similar to other geophysical simulators, the model has highly uncertain parameters that need to be calibrated to generate realistic simulations \citep{wang2019parallel}. Our particular interest is in calibrating the following five parameters: Manning’s roughness coefficient for channel type \#5 ($\theta_1$), overland roughness control factor ($\theta_2$), deep drainage ($\theta_3$), infiltration-scaling parameter ($\theta_4$), and saturated soil lateral conductivity ($\theta_5$). The first two parameters usually affect the shape and timing of peaks in hydrographs, which are time series plots showing water discharge at a certain location of a river. The remaining parameters determine the total water volume in hydrographs. The range of each parameter is rescaled to a unit interval $[0,1]$ for ease of exposition. We calibrate the model based on the observed hydrograph at Iowa River at Wapello, IA
(USGS site ID\#05465500), a time series of water discharge with 480 time steps \citep{bhatnagar2022computer}. We use a uniform prior on the unit interval $[0,1]$ for $\theta_1,\dots,\theta_5$ and slightly informative priors for the discrepancy parameters as in \cite{bhatnagar2022computer}. The prior specifications are listed in Table S3 in the Supplementary Material.

One challenge that distinguishes this calibration problem from the previous two examples is the computational cost of the WRF-Hydro model. Unlike the SIR models used previously, the WRF-Hydro model consists of numerous partial differential equations and requires over 30 minutes to obtain one simulation run, even with a high-performance workstation with many cores. Here, we employ a Gaussian process (GP) emulator \cite{sacks1989design,kennedy2001bayesian} as the forward model $A(\boldsymbol{\theta})$, serving as a surrogate for the original WRF-Hydro model output. The GP emulator is constructed based on the WRF-Hydro runs solved for the 400 pre-defined parameter settings $\boldsymbol{\theta}^{(1)},\dots,\boldsymbol{\theta}^{(400)}$ determined by Latin hypercube design \citep{bhatnagar2022computer}, and is designed to predict the outcomes of the WRF-Hydro model at any parameter setting $\mathbf{\theta}$. For the covariance function, we utilize the standard separable covariance for the parameter space and time domain \citep[see, e.g.,][]{olson2013mathematical,RJ-2018-049}. Figure S4 depicts the model runs used to construct the emulator $A(\boldsymbol{\theta})$ and the observational data ($\mathbf{D}$) used for calibration. The data-model discrepancy $\boldsymbol{\delta}$ is also represented as another Gaussian process with an exponential covariance to account for structural differences between the model output and observational data.

This example illustrates how our method operates when the original model is replaced with a GP emulator, which is the standard approach when repeated evaluations of the computer model are not computationally feasible. Here, we anticipate that the shape of the posterior distribution will be complex, and even the adaptive MCMC may encounter difficulties in inference. This complexity arises from the notable data-model discrepancy, which is highly common for geophysical models \citep[see, e.g.,][]{chang2013fast,chang2015binary,salter2019uncertainty}. Nonidentifiability can make it challenging to distinguish the effects from $A(\boldsymbol{\theta})$ and $\bm{\delta}$, which is a well-known problem in the computer model calibration literature \citep{chang2013fast,brynjarsdottir2014learning,salter2019uncertainty,gu2019jointly}.



For ATVI, we use 20 layers (10 for $t^{(1)}=1.7$ and 10 for $t^{(2)}=1$) with the learning rate of $8 \times 10^{-3}$. Similarly, we use 20 layers for VI (NF) with the learning rate of $3 \times 10^{-3}$. For VI (copula), the learning rate is set to be $6 \times 10^{-3}$. For the adaptive MCMC algorithm, we run it for 500,000 iterations with 5 different random initial starting points. We report the results from the chain that achieves the highest (unnormalized) log maximum a posteriori probability (log MAP). Here, we need a large number of iterations for the adaptive MCMC to ensure a well-mixed chain, due to the complications explained above.

\begin{figure}[ht]
\begin{center}
    \begin{tabular}{c c}
        \textbf{ATVI} & \textbf{VI (NF)} \\
        \includegraphics[width=0.45\columnwidth]{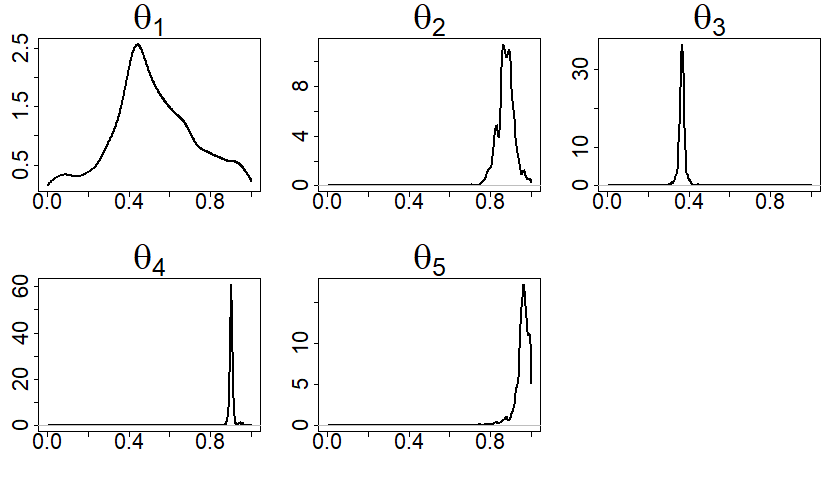} & \includegraphics[width=0.45\columnwidth]{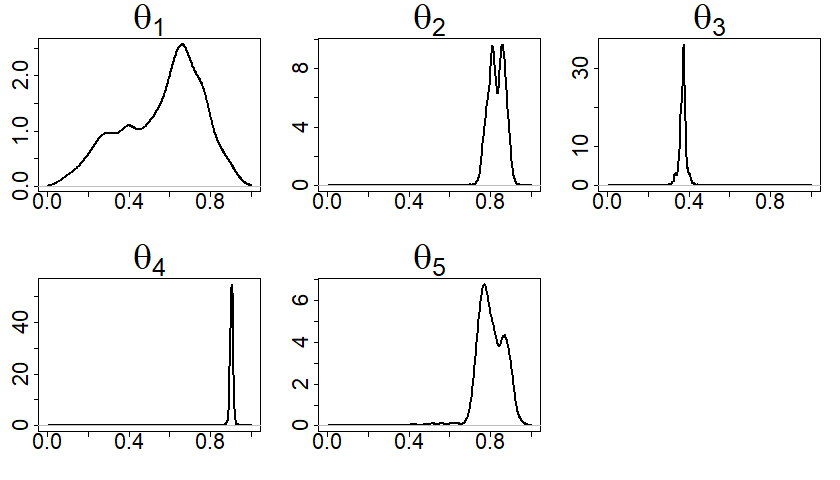} \\
        \textbf{VI (copula)}& \textbf{MCMC} \\ 
        \includegraphics[width=0.45\columnwidth]{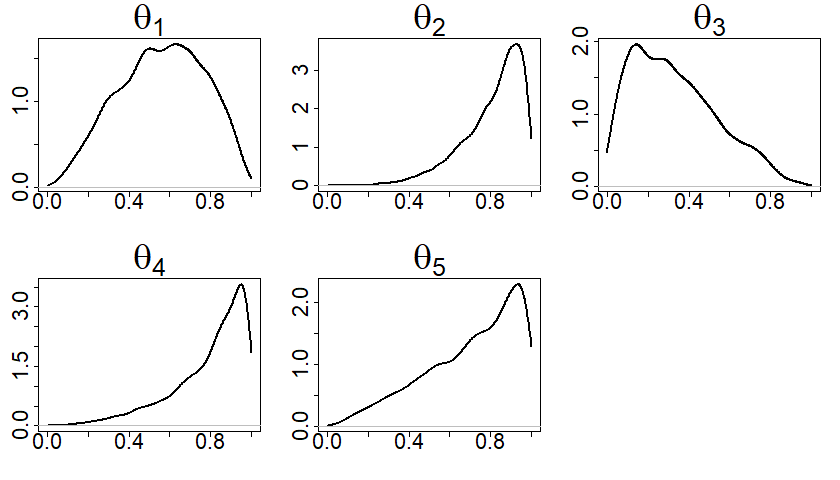}
         & \includegraphics[width=0.45\columnwidth]{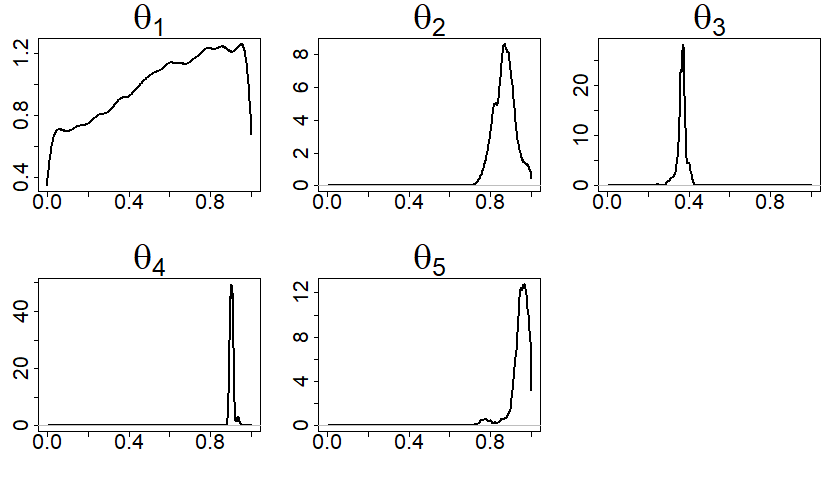} \\ 
    \end{tabular}
\end{center}
\caption[]{The marginal posterior densities of model parameters in the WRF-Hydro model, estimated by all compared models.}
\label{WRF_density}
\end{figure}

The posterior densities for the five input parameters ($\theta_1, \cdots, \theta_5$) in Figure \ref{WRF_density} and the HPD intervals in Table \ref{table:table_koh_result} indicate that ATVI yields results most similar to those from the adaptive MCMC. Figure \ref{WRF_density} illustrates that VI (NF) clearly encounters boundary issues for $\theta_2$ and $\theta_5$, while VI (copula) appears to lead to much higher estimation uncertainties for all parameters except for $\theta_1$ compared to the other methods. Table \ref{WRF_density} also presents the log MAP for different methods. Here, the adaptive MCMC achieves the highest log MAP, and ATVI achieves a close value. The other two VI methods yield notably lower log MAP values.

We also investigate how the different posterior densities obtained by the compared methods affect the prediction performance of the fitted model, which can be represented as the sum of the emulator and discrepancy terms, $A(\boldsymbol{\theta})+\boldsymbol{\delta}$. Since $A(\boldsymbol{\theta})+\boldsymbol{\delta}$ is a multivariate normal random vector that represents our predictive process for $\mathbf{D}$, we assess its interpolation performance by (i) treating a part of $\mathbf{D}$ as the `observed values' and (ii) predicting the rest of $\mathbf{D}$, the `unobserved values', based on the observed values as well as the mean and covariance of $A(\boldsymbol{\theta})+\boldsymbol{\delta}$. Similar to standard Gaussian process prediction, we utilize the conditional normal distribution of the unobserved values given the observed values to determine the predictive mean and the 95\% prediction interval. In our experiment, we set 48 equally spaced time steps as our observed values and attempt to predict the remaining 432 values to compute the performance metrics. The results are summarized in Table \ref{table:table_koh_result}.

In terms of MSPE, ATVI, VI (NF), and the adaptive MCMC perform similarly, with ATVI slightly outperforming the others. Regarding the prediction interval, the results indicate that ATVI leads to slightly shorter average prediction intervals (1976.18) compared to the adaptive MCMC (2251.98) while achieving closer empirical coverage (97.0\%) than the adaptive MCMC (98\%).  ATVI outperforms the VI (NF) and VI (copula) as well; hence, our modification using boundary surjection and temperature annealing is helpful for better inference results in this example. Additionally, the computation time for ATVI (2.4 hours) is significantly shorter than that for the adaptive MCMC (174.5 hours). 
Figure S5 illustrates the prediction intervals from all compared methods, indicating that the prediction intervals from ATVI are the narrowest among the compared methods.

\begin{table}[htbp]
{
\centering
\begin{tabular}{ccccc}
\hline
& ATVI & VI (NF) & VI (copula) & MCMC \\
\hline
$\theta_1$ & 0.496 & 0.610 &0.570 & 0.592\\
95\% HPD & (0.156, 1.000) & (0.193, 0.897) & (0.187, 0.942) &(0.076, 1.000) \\
\hline
$\theta_2$ & 0.874 & 0.829 & 0.847 & 0.869\\
95\% HPD & (0.789, 0.962) & (0.757, 0.896) & (0.512, 0.995) & (0.772, 0.981) \\
\hline
$\theta_3$  & 0.366 & 0.372 & 0.301 & 0.369\\
95\% HPD & (0.330, 0.400) & (0.329, 0.410) & (0.009, 0.736)& (0.317, 0.417)\\
\hline
$\theta_4$  & 0.900 & 0.902 & 0.848 & 0.902\\
95\% HPD & (0.877, 0.953) & (0.885, 0.921) & (0.421, 0.999)  & (0.883, 0.918)\\
\hline
$\theta_5$  & 0.960 & 0.795& 0.753 & 0.953\\
95\% HPD & (0.878, 1.000) & (0.700, 0.919) & (0.264, 1.000) & (0.863, 1.000)\\
\hline
\hline
Comp. Time (hours)& 2.4 & 1.2 & 1.4 & 174.5 \\
Data Coverage & 97.0 & 97.4 &  99.0 & 98.1 \\
Data AIL & 1976.18 & 2281.04 & 2793.91 &  2251.98\\  
Data MSPE  &376945& 480777 & 683139& 458710 \\
log MAP & -3573.2 & -3577 & -3583.6  & -3572.1 \\
\hline
\end{tabular}
\caption{Inference results for the WRF-Hydro model. For all parameters, mode of estimates and 95\% HPD interval are reported. The data coverage, AIL, and MSPE are based on prediction analysis based on the conditional distribution given by $A(\boldsymbol{\theta})+\boldsymbol{\delta}$ as described in Section \ref{section:WRF}. The computing time is in hours.}
\label{table:table_koh_result}
}
\end{table}

\vspace{-20pt}

\section{Discussion}

We propose a fast and flexible VI algorithm for computer model calibration. To construct a variational distribution, we utilize normalizing flows that do not require specific parametric models. Furthermore, we embed surjective transformation and temperature annealing within our framework to alleviate inferential challenges in a naive VI algorithm, boundary and local optima problems when applied to complex computer models. In the two SIR model calibration examples, ATVI provides comparable inference results as the baseline Bayes approach (adaptive MCMC), while it is computationally much faster. In the WRF-Hydro model calibration example, our method led to a better inference result than the baseline VI methods achieving log MAP closer to the adaptive MCMC. The ideas and method proposed in this paper can generally be applicable to other calibration tasks such as biochemical models \citep{hwang2021bayesian}. 

There can be several improvements for ATVI from an algorithmic perspective. We observed that sequentially annealed posteriors were not sensitive depending on the temperature ladder once we set the first temperature to be high enough. However, developing an adaptive temperature schedule in ATVI would result in a more automatic algorithm. Furthermore, since there are many different divergences with various trade-offs \citep{regli2018alpha}, selecting an optimal divergence for a given variational inference problem remains an open question. 

Another possible direction is to utilize other types of variational neural networks for calibration. For example, one might be able to build an algorithm similar to variational autoencoder \citep{kingma2014autoencoding}, with an encoder turning the observed data into latent variables and a decoder transforming the latent variables into estimated parameter values. Such development will pose interesting theoretical challenges, including the formation of a suitable loss function and defining flexible latent distributions.

\vspace{-5pt}
\section*{Acknowledgement}
\vspace{-5pt}  
Jaewoo Park was partially supported by the National Research Foundation of Korea (2020R1C1C1A0100386814, RS-2023-00217705) and ICAN (ICT Challenge and Advanced Network of HRD) support program (RS-2023-00259934) supervised by the IITP (Institute for Information \& Communications Technology Planning \& Evaluation). Won Chang was partially supported by the University of Cincinnati TAFT Research Center. 

\vspace{-5pt}
\section*{Supplementary Material}
\vspace{-5pt}
Supplementary material available online contains proofs, prior settings, model details, additional results, and an example trace plot of loss. The source code can be downloaded from \url{https://github.com/CrawlingKiming/AdVI}.

\bibliography{Reference.bib}



\section*{Supplementary material}

# Supplementary Material for Fast Computer Model Calibration using Annealed and Transformed Variational Inference

Dongkyu Derek Cho, Won Chang, and Jaewoo Park

March 5, 2024

## S1 Prior Settings

| Parameter | Definition | Prior |
|---|---|---|
| $\beta$ | Transmission rate per day | Uniform (0, 3) |
| $\gamma$ | Recovery rate per day | Uniform (0, 3) |
| $S_0$ | Initial number of susceptibles | Uniform (37, 100) |

Table S1: Prior distributions for the SIR model parameters.

| Parameter | Definition | Prior |
|---|---|---|
| $w$ | Amplitude of transmission rate | Uniform (0, 1) |
| $\phi$ | Offset of the seasonal variation | Uniform (2, 8.28) |
| $\nu$ | Dispersion rate | Uniform (0.5, 1) |
| $\rho$ | Reporting rate | Uniform (0, 1) |
| $\beta_{0i}$ | Base rate parameter for age class $i$, $i \in \{1, 2, ..., 6\}$ | Uniform (0, 4) |

Table S2: Prior distributions for the MSIR model parameters

| Parameter | Definition | Prior |
|---|---|---|
| $\theta_i$ | Scaled WRF-Hydro model parameter $i$, $i \in \{1, 2, 3, 4, 5\}$ | Uniform (0, 1) |
| $\delta_{\text{range}}$ | Range parameter in GP for $\boldsymbol{\delta}$ | Gamma (2, 150) |
| $\delta_{\text{sill}}$ | Sill parameter in GP for $\boldsymbol{\delta}$ | Inverse-Gamma (2, 10500) |
| $\delta_{\text{nugget}}$ | Nugget parameter in GP for $\boldsymbol{\delta}$ | Inverse-Gamma (2, 16.5) |

Table S3: Prior distributions used for WRF-Hydro calibration

# S2 Model Details

## S2.1 MSIR Model

Park et al. (2017) assumed the birth rate $\mu(t)$ changes over time. Park et al. (2017) considered the mean weekly birth rate as $\bar{\mu} = 1/(5 \times 52)$ with variation in monthly birth rate (Table S4). Seasonal variation in the birth rate in Niger is estimated from Demographic and Health Surveys (Dorélien, 2013).

| Jan | Feb | Mar | Apr | May | Jun | Jul | Aug | Sep | Oct | Nov | Dec |
|---|---|---|---|---|---|---|---|---|---|---|---|
| -0.17 | 0.01 | 0.03 | 0.25 | 0.12 | 0.03 | -0.01 | 0.09 | 0.01 | 0.13 | -0.31 | -0.17 |

Table S4: Seasonal variation in the birth rate.

As in Park et al. (2017), we use a contact matrix

$$C = \begin{bmatrix} 1 & 1 & 1 & 1 & 1 & 1 \\ 1 & 1 & 1 & 1 & 1 & 1 \\ 1 & 1 & 1 & 1 & 1 & 1 \\ 3 & 3 & 3 & 1 & 1 & 1 \\ 6 & 6 & 6 & 2 & 1 & 1 \\ 18 & 18 & 18 & 6 & 3 & 1 \end{bmatrix}$$

Here, $C_{ij}$ denotes the frequency of contact between age class $i$ and $j$, satisfying $f_i C_{ij} = f_j C_{ji}$, where $f_i$ is the fraction of the age group $i$ of the population. The model using such a contact matrix assumes that the contact between age groups is homogeneous. Furthermore, Park et al. (2017) only considered transmission between children and assumed the population of children is closed.

## S2.2 Rational Quadratic Flows Details

The transformed values in (7) induce a set of knot points $\{u^{(m)}, v^{(m)}\}_{m=0}^{M+1}$ and a set of slope values $\{s^{(m)}\}_{m=1}^{M}$ that are defined by

$$u^{(m+1),(i)} = u^{(m),(i)} + w_m^{(i)}, m = 1, ..., M \tag{S1}$$

$$v^{(m+1),(i)} = v^{(m),(i)} + e_m^{(i)}, m = 1, ..., M \tag{S2}$$

$$s^{(m),(i)} = (v^{(m+1),(i)} - v^{(m),(i)})/w_m^{(i)}, m = 1, ..., M \tag{S3}$$

with $u^{(0),(i)}, v^{(0),(i)} = -\infty, u^{(1),(i)}, v^{(1),(i)} = -2, u^{(M),(i)}, v^{(M),(i)} = 2, u^{(M+1),(i)}, v^{(M+1),(i)} = \infty$. For brevity, we suppress the superscript $(i)$ for $\{u^{(m),(i)}, v^{(m),(i)}\}_{m=0}^{M+1}$ and denote them by $\{u^{(m)}, v^{(m)}\}_{m=0}^{M+1}$. Let $x$ be an any point in $m$th knot point, $x \in [u^{(m)}, u^{(m+1)}]$ and define $r_x := (x - u^{(m)})/w^{(m)}$. For $m = 1, \ldots, M-1$, the rational-quadratic spline function $\zeta^{(m)}$ is given by

$$\zeta^{(m)}(x) = \frac{\zeta_1^{(m)}(r_x)}{\zeta_2^{(m)}(r_x)}, \tag{S4}$$

where

$$\zeta_1^{(m)}(r_x) = s^{(m)} v^{(m+1)} r_x^2 + [v^{(m)} \rho^{(m+1)} + v^{(m+1)} \rho^{(m)}] r_x (1 - r_x) + s^{(m)} v^{(m)} (1 - r_x)^2$$

$$\zeta_2^{(m)}(r_x) = s^{(m)} r_x^2 + [\rho^{(m+1)} + \rho^{(m)}] r_x (1 - r_x) + s^{(m)} (1 - r_x)^2.$$

For $m = \{0, M\}$, we let $\zeta^{(m)}$ be an identity function, $\zeta^{(0)}(x) = \zeta^{(M)}(x) = x$. As a result, $\zeta^{(m)}$ is given by

$$y_i = \zeta^{(m)}(x_i) = \begin{cases} x_i, & \text{if } x_i \in [u^{(0)}, u^{(1)}], \\ \zeta^{(m)}(x_i), & \text{if } x_i \in [u^{(m)}, u^{(m+1)}], \\ x_i, & \text{if } x_i \in [u^{(M)}, u^{(M+1)}]. \end{cases} \quad (S5)$$

The Jacobian term for the transformation given by the above formalation can be easily computed. We start with the following equation:

$$\zeta_2^{(m)}(r_x) \left[ \frac{d\zeta_1^{(m)}(r_x)}{dr_x} \right] - \zeta_1^{(m)}(r_x) \left[ \frac{d\zeta_2^{(m)}(r_x)}{dr_x} \right] = w^{(m)} (s^{(m)})^2 [\rho^{(m+1)} r_x^2 + 2s^{(m)} r_x (1 - r_x) + \rho^{(m)} (1 - r_x)^2]. \quad (S6)$$

The derivative of an equation is found using the quotient rule,

$$\begin{aligned} \frac{\partial}{\partial x} [\zeta^{(m)}(x)] &= \frac{\partial}{\partial x} \left[ \frac{\zeta_1^{(m)}(r_x)}{\zeta_2^{(m)}(r_x)} \right] \\ &= \frac{\zeta_2^{(m)}(r_x) \frac{\partial}{\partial r_x} [\zeta_1^{(m)}(r_x)] - \zeta_1^{(m)}(r_x) \frac{\partial}{\partial r_x} [\zeta^{(m)}(r_x)]}{w^{(m)} [\zeta_2^{(m)}(r_x))]^2} \\ &= \frac{w^{(m)} (s^{(m)})^2 [\rho^{(m+1)} r_x^2 + 2s^{(m)} r_x (1 - r_x) + \rho^{(m)} (1 - r_x)^2]}{s^{(m)} + [\rho^{(m+1)} + \rho^{(m)} - 2s^{(m)}] r_x (1 - r_x)]^2} \end{aligned} \quad (S7)$$

which becomes the Jacobian term for a single rational quadratic spline layer. The last line in (S7) results from (S6). Therefore, regardless of the choices of the function $\sigma_i(\cdot)$, the determinant can be easily computed. By generalizing the above, we can construct a deep generative model with multiple rational quadratic spline layers, resulting in a highly flexible function with low computational cost.

# S3 Proof of the Main Results

## S3.1 Proposition 1

**Proof** Consider the logarithm of the target density $\log p_{\boldsymbol{\xi}}(\boldsymbol{\xi}|\boldsymbol{D}) = \log p_{\boldsymbol{\theta}}(\boldsymbol{\theta}|\boldsymbol{D}) + \mathcal{V}(\boldsymbol{\theta}, \boldsymbol{\xi})$, where

$$\mathcal{V}(\boldsymbol{\theta}, \boldsymbol{\xi}) = \sum_{i=1}^{d} \log w_{s_i}(s_i|\theta_i) = \sum_{i=1}^{d} \delta(s_i = g_2(\xi_i)) \log w_{s_i}(s_i|\theta_i)$$

$$= \sum_{i=1}^{d} \left[ \delta(0 = g_2(\xi_i)) \log(1 - u_i(\theta_i)) + \delta(1 = g_2(\xi_i)) \log(u_i(\theta_i)) \right].$$

The second equality holds because we set $g_2(\xi_i) = s_i$. As both $p_{\boldsymbol{\theta}}(\boldsymbol{\theta}|\boldsymbol{D})$ and $u_i(\theta_i)$ are continuous functions, we only need to show that $\mathcal{V}(\boldsymbol{\theta}, \boldsymbol{\xi})$ is continuous at the parameter boundary (i.e., $\xi_i \to a_i$ for some $i$). Due to the Condition 2, $\forall i, u_i(a_i) = 1 - u_i(a_i) = 0.5$ and thus $p_{\boldsymbol{\xi}}(\boldsymbol{\xi}|\boldsymbol{D})$ is continuous.

Consider a point $\boldsymbol{k} \in \prod_{i=1}^{d}(-\infty, a_i - r) \cap \mathcal{B}^{-1}(\boldsymbol{\Theta})$. Then we have the following result.

$$\mathcal{V}(\boldsymbol{\theta} = 2\boldsymbol{a} - \boldsymbol{k}, \boldsymbol{\xi} = \boldsymbol{k}) = \sum_{i=1}^{d} \left[ \delta(0 = g_2(k_i)) \log(1 - u_i(2a_i - k_i)) \right] = \sum_{i=1}^{d} \left[ \log(1 - u_i(2a_i - k_i)) \right]$$

The equalities hold due to the fact that $\mathcal{B}(\boldsymbol{k}) = 2\boldsymbol{a} - \boldsymbol{k}$ and $\forall i, g_2(k_i) = 0$. Then we have

$$p_{\boldsymbol{\xi}}(\boldsymbol{k}|\boldsymbol{D}) = p_{\boldsymbol{\xi}}(\boldsymbol{k}|\boldsymbol{D}) \prod_{i=1}^{d}(1 - u_i(2a_i - k_i))$$

$$\leq M \prod_{i=1}^{d} \sqrt[d]{\epsilon/M} = \epsilon$$

from Condition 3. Hence, $p_{\boldsymbol{\xi}}(\boldsymbol{k}|\boldsymbol{D}) \leq \epsilon$.

The Condition 4 is equivalent to

$$\frac{\partial u_i(c_i)}{\partial c_i}(1 - u_i(c_i))^{-1} \geq \frac{\partial p_{\boldsymbol{\theta}}(\boldsymbol{c}|\boldsymbol{D})}{\partial c_i}(p_{\boldsymbol{\theta}}(\boldsymbol{c}|\boldsymbol{D}))^{-1}$$

$$\Leftrightarrow \frac{\partial u_i(c_i)}{\partial c_i} p_{\boldsymbol{\theta}}(\boldsymbol{c}|\boldsymbol{D}) - (1 - u_i(c_i))\frac{\partial p_{\boldsymbol{\theta}}(\boldsymbol{c}|\boldsymbol{D})}{\partial c_i} \geq 0$$

$$\Leftrightarrow \prod_{j \neq i}(1 - u_j(c_j))\frac{\partial u_i(c_i)}{\partial c_i} p_{\boldsymbol{\theta}}(\boldsymbol{c}|\boldsymbol{D}) - \prod_{j}(1 - u_j(c_j))\frac{\partial p_{\boldsymbol{\theta}}(\boldsymbol{c}|\boldsymbol{D})}{\partial c_i} \geq 0$$

$$\Leftrightarrow \frac{\partial \left(\prod_{j}(1 - u_j(c_j))p_{\boldsymbol{\theta}}(\boldsymbol{c}|\boldsymbol{D})\right)}{\partial(-c_i)} \geq 0$$

$$\Leftrightarrow \frac{\partial p_{\boldsymbol{\xi}}(2\boldsymbol{a} - \boldsymbol{c}|\boldsymbol{D})}{\partial(-c_i)} \geq 0,$$

where the last equivalence comes from the fact that $\mathcal{B}(2\boldsymbol{a} - \boldsymbol{c}) = \boldsymbol{c}$ and $\forall i, g_2(2a_i - c_i) = 0$. Therefore, $p_{\boldsymbol{\xi}}(\boldsymbol{\xi}|\boldsymbol{D})$ is a monotone increasing function within $\prod_{i=1}^{d}[a_i - r, a_i] = \mathcal{Q}_{r-}(\boldsymbol{a})$. For the transformed approximated distribution, we have $\log q_{\boldsymbol{\theta}}(\boldsymbol{\theta}; \boldsymbol{\phi}) = \log q_{\boldsymbol{\xi}}(\boldsymbol{\xi}; \boldsymbol{\phi}) - \mathcal{V}(\boldsymbol{\theta}, \boldsymbol{\xi})$. Since the Proposition 1 ensures $u_i(\theta_i)$ and $1 - u_i(\theta_i)$ are both in class of $C^\infty$, $\mathcal{V}(\boldsymbol{\theta}, \boldsymbol{\xi}) \in C^\infty(\mathcal{Q}_{r+}(\boldsymbol{a}))$. Thus, $q_{\boldsymbol{\theta}}(\boldsymbol{\theta}; \boldsymbol{\phi}) \in C^m(\mathcal{Q}_{r+}(\boldsymbol{a}))$ □

## S3.2 Remark 2

**Proof** Obviously, $\varsigma_i(x; a_i, B)$ satisfies the first and second conditions. Further, note that $\varsigma_i(B; a_i, x)$ is an strictly increasing function with respect to $B$, such that $\varsigma_i(B; a_i, x) : (-\infty, \infty) \to (0, 1)$ for $x \in [a_i, \infty)$. For some small $r \in (0, \frac{b_i - a_i}{2})$, we can find a $B$ that is large enough to satisfy the third condition. Now, note a derivative of the logistic function can be written as follows:

$$\frac{\partial \varsigma_i(x; A, 1)}{\partial x} = \varsigma_i(x; A, 1)\left(1 - \varsigma_i(x; A, 1)\right).$$

Therefore, $\frac{\partial \varsigma_i(x; A, B)}{\partial x} = B \varsigma_i(x; A, B)\left(1 - \varsigma_i(x; A, B)\right)$. The left term of Condition 4 in Proposition 2 now becomes identical to Section 4.2 (18) because

$$\frac{\partial \varsigma_i(c_i; a_i, B)}{\partial c_i}(1 - \varsigma_i(c_i; a_i, B))^{-1} = \frac{B \varsigma_i(c_i; a_i, B)\left(1 - \varsigma_i(c_i; a_i, B)\right)}{1 - \varsigma_i(c_i; a_i, B)}$$

$$= B \varsigma_i(c_i; a_i, B)$$

□

# S4 Figures for Section 5.1

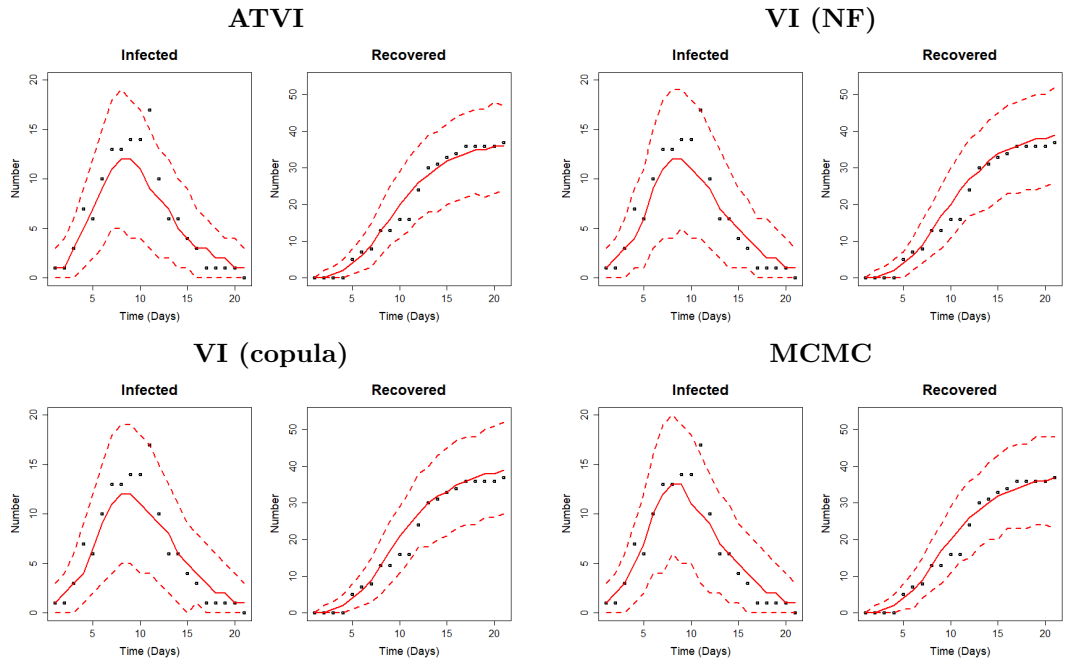

Figure S1: The true number of infected and recovered population (black dots) from the basic SIR model. Solid and dashed lines indicate posterior mean and corresponding 95% HPD interval, respectively.

# S5 Figures for Section 5.2

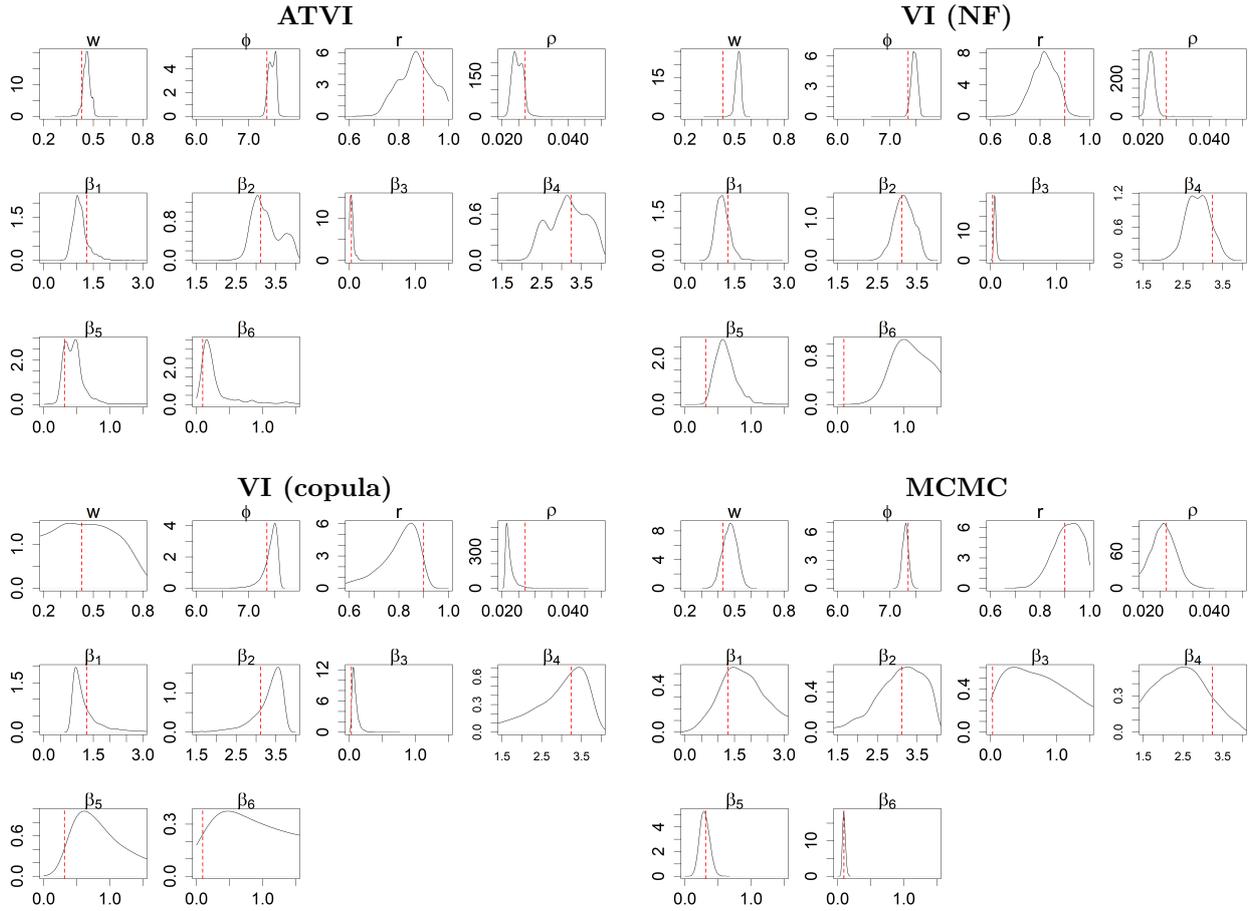

Figure S2: Posterior densities of the parameters in MSIR model from an example iteration out of 50 repeated simulations. For all figures, red lines indicate the true parameter values. For some figures the x-axes are truncated so that some densities do not appear too narrow.

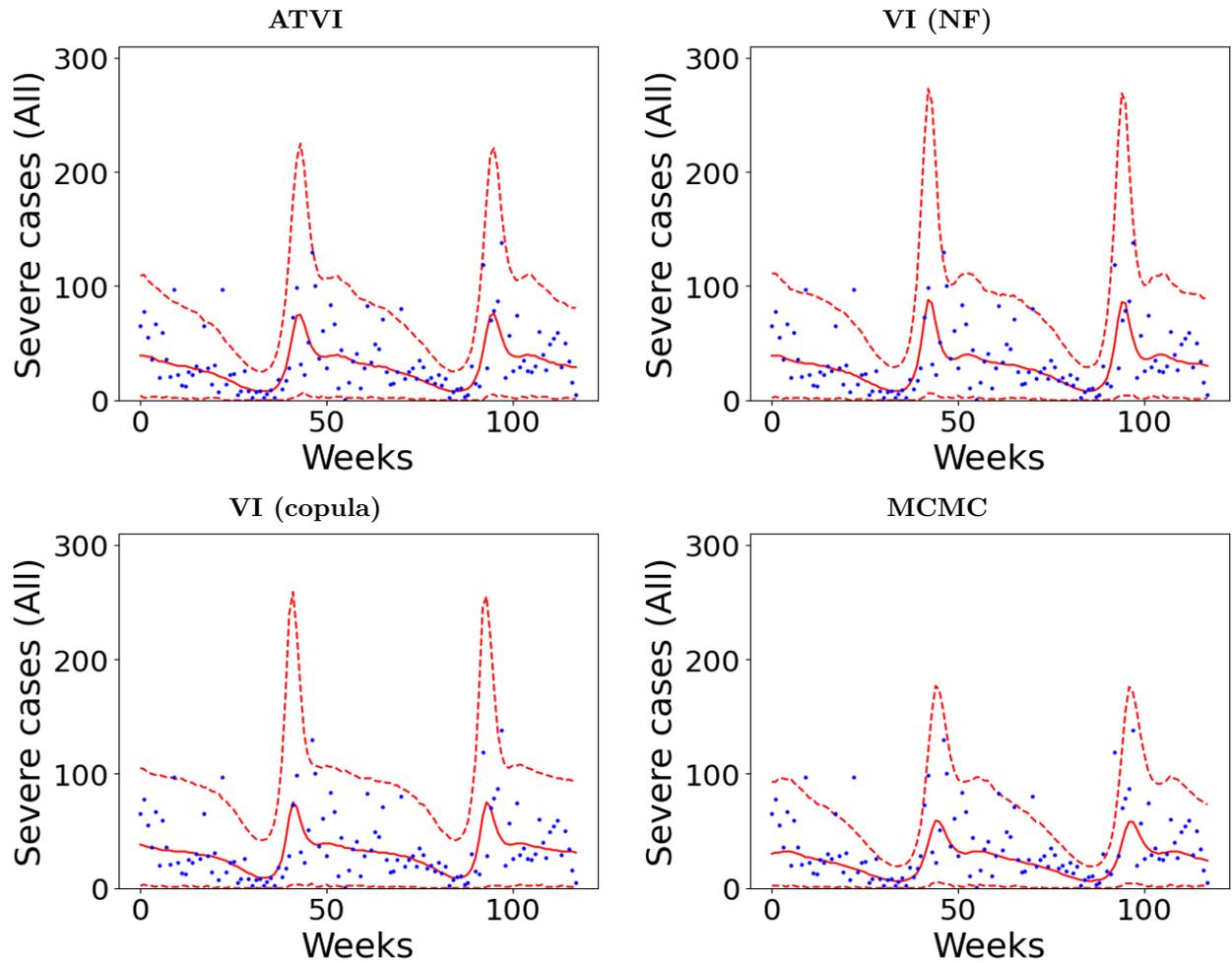

Figure S3: The observed number of cases (blue dots) and model projection (red lines). For all figures, solid and dashed lines indicate posterior mean and corresponding 95% HPD interval, respectively.

8

# S6   Figures for Section 5.3

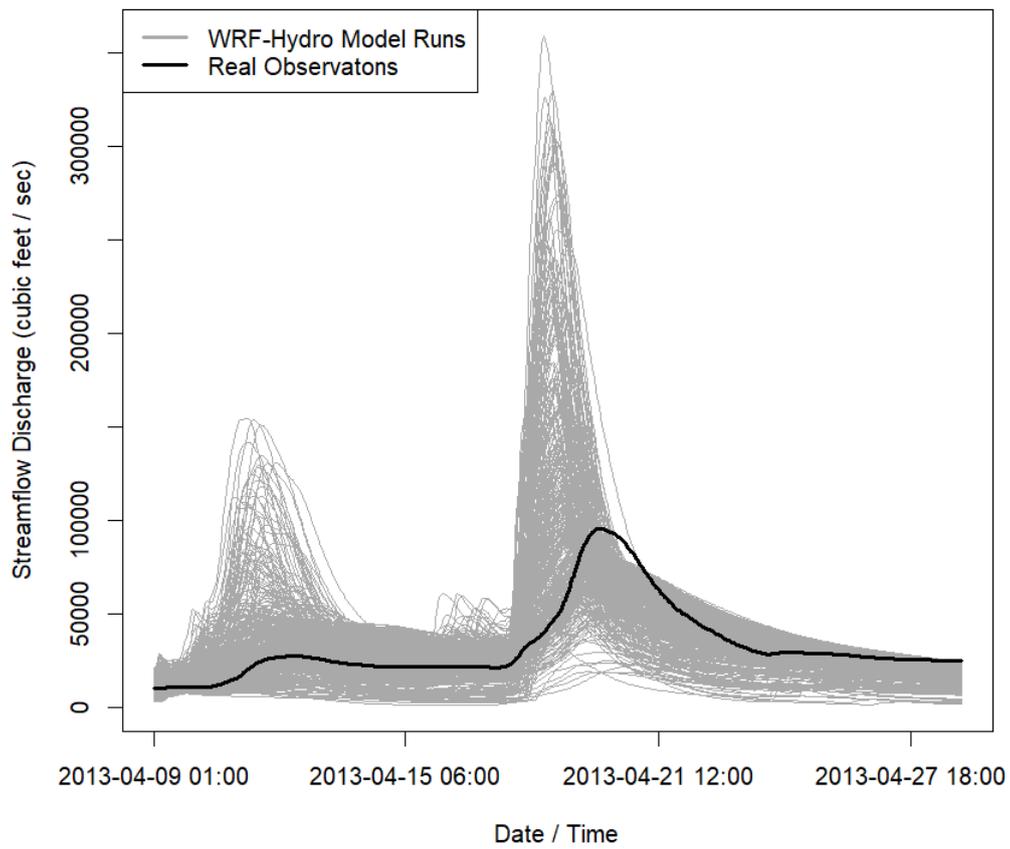

Figure S4: WRF-Hydro model runs (grey lines) and observed data (black line) at 480 time series for the stream flow charge from April 9th to 28th in 2013 at Iowa River at Wapello, IA (USGS site ID#05465500). The Figure is adopted from Bhatnagar et al. (2022) with minor modifications.

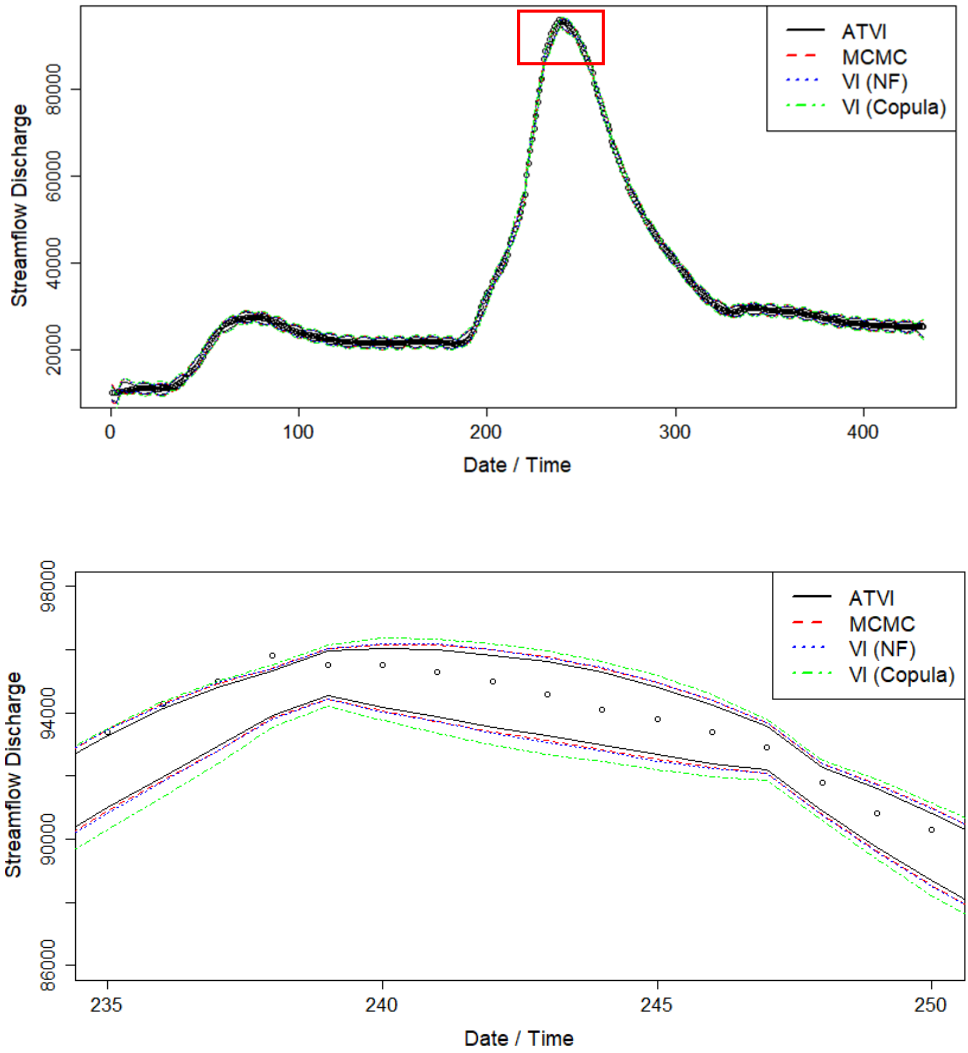

Figure S5: (Top) Results from prediction analysis based on the conditional distribution given by $A(\boldsymbol{\theta}) + \boldsymbol{\delta}$ as described in Section 5.3 (Bottom) A zoomed up figure focuing on the red-boxed area in the top panel.

# S7 Example Loss Curves for VI Methods

The figure below shows the loss curves from all VI methods in the SIR example (Section 5.1). The loss curves in the other examples show similar patterns.

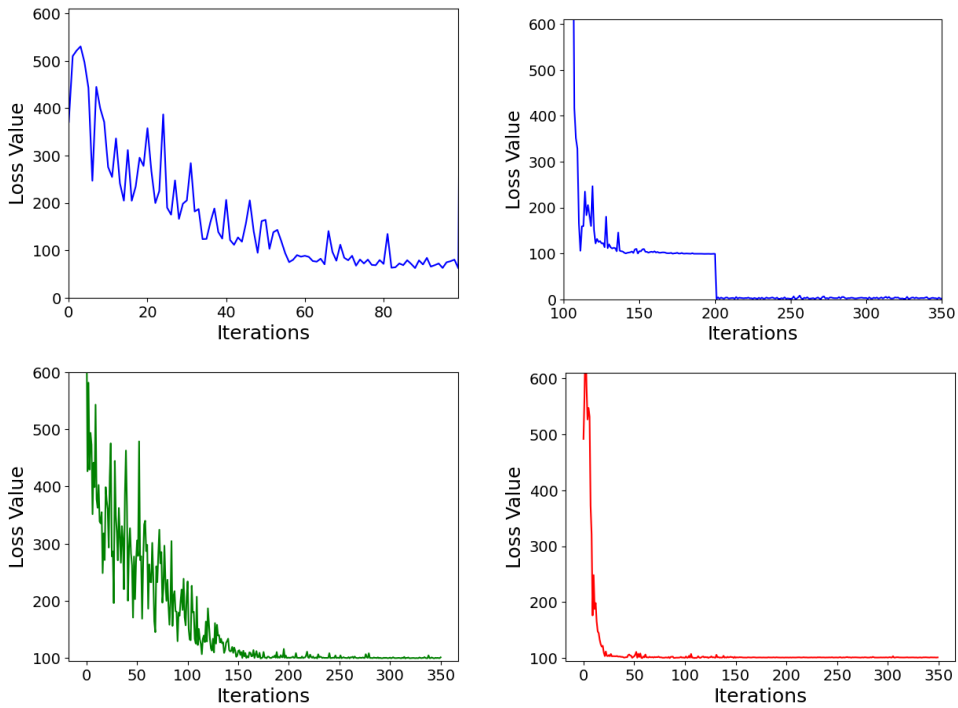

Figure S6: (Top left) The loss curve for ATVI with the first temperature ($t^{(1)} = 3$) (Top right) The loss curve for ATVI with the second temperature ($t^{(1)} = 1$). The weighted adjusted fine tuning explained in Section 4.3 is applied after the 200th iteration (Bottom left) The loss curve for VI (NF) (Bottom right) The loss curve for VI (copula)



\end{document}